\documentclass[journal]{IEEEtran}
\usepackage{threeparttable}

\usepackage[utf8]{inputenc}
\usepackage{textcomp}
\usepackage{amsmath}
\usepackage{amssymb}
\usepackage{amsfonts}

\usepackage{cite}           
\usepackage{graphicx}       
\usepackage{float}          

\usepackage{hyperref}       
\usepackage{cleveref}       
\usepackage{hyperref}       
\usepackage{cite}
\usepackage{algorithmic}

\usepackage{xcolor}
\usepackage{newunicodechar}
\newunicodechar{−}{\ensuremath{-}}
\usepackage{booktabs}
\usepackage{url}
\usepackage{accessibility}
\def\BibTeX{{\rm B\kern-.05em{\sc i\kern-.025em b}\kern-.08em
    T\kern-.1667em\lower.7ex\hbox{E}\kern-.125emX}}

\usepackage{cite}
\usepackage{amsmath,amssymb,amsfonts}
\usepackage{algorithmic}
\usepackage{graphicx}
\usepackage{textcomp}
\usepackage{tabularx}
\usepackage{xcolor}
\usepackage{booktabs}
\def\BibTeX{{\rm B\kern-.05em{\sc i\kern-.025em b}\kern-.08em
    T\kern-.1667em\lower.7ex\hbox{E}\kern-.125emX}}

\begin{document}
%
\title{A Conductance-Based Amygdala Model of Threat Processing
in Anxiety and Depression}
%
%
%
%

\author{Malik~Faizan,
        P.J.~White,
        and~Indrakshi Dey,~\IEEEmembership{Senior Member,~IEEE}
\thanks{Malik~Faizan and Indrakshi Dey are with the Walton Institute for Information and Communication System Science, Southeast Technological University, Ireland.}%
\thanks{P.J.~White is with the Department of Humanities, Southeast Technological University, Ireland.}
\thanks{***This work has been submitted to the IEEE for possible publication. Copyright may be transferred without notice, after which this version may no longer be accessible.}}

\maketitle

\begin{abstract}
Anxiety and depressive disorders are increasingly viewed as dysregulations along continuous stress-regulatory dimensions. However, existing computational approaches seldom connect interpretable circuit-level mechanisms to autonomic physiology. \textbf{Methods:} This study develops a mechanistic framework that links amygdala dysregulation to cardiovascular stress responses for digital phenotyping and clinical interpretation.  We formulated a compact, nine-equation, conductance-based model of the amygdala–hypothalamus–cardiovascular pathway. The framework extends Hodgkin–Huxley-formalism with three clinically grounded modulators: coping capacity, perceived stress load, and prefrontal regulatory strength. A slow, history-dependent internal state, adaptive thresholding, graded threat acknowledgment, and baroreflex-coupled hypothalamic integration were used to generate heart rate and blood pressure trajectories. \textbf{Results:} Distinct strong, moderate, and weak regulatory regimes emerged as stable operating states of a single closed-loop system. Robust analyses showed that stochastic variability and parameter perturbation preserved regime separation, while ablation studies identified the adaptive threshold as the principal mechanism driving quantitative regime separation. Simulated cardiovascular responses remained within reported stress-physiology ranges. Furthermore, external evaluation across three independent datasets supported robust agreement with real-world, stress-related autonomic patterns. \textbf{Conclusion:} A compact, mechanistic model can jointly link psychometric modulators, amygdala excitability, and downstream cardiovascular output within a single, interpretable framework. \textbf{Significance:} This work provides a computationally tractable basis for mechanism-informed digital phenotyping, patient-specific stress monitoring, and future digital-twin approaches for mental-health decision support
\end{abstract}


%
\IEEEpeerreviewmaketitle

\section{Introduction}
\IEEEPARstart{A}{nxiety} and major depressive disorders are among the most prevalent mental health conditions globally, affecting nearly one billion individuals and serving as a leading cause of disability \cite{gbd2019}. Both disorders share a clinical profile characterized by excessive threat appraisal, diminished recovery from acute stressors, and chronic sympathetic activation \cite{vaccarino2024,cheng2022,tomasi2024}. Rather than discrete categories, these phenomena are increasingly viewed as quantitative displacements along continuous regulatory dimensions \cite{huys2021}. A perspective formalized by transdiagnostic and dynamical‑systems frameworks in computational psychiatry that treat neural‑circuit dysfunction as the primary descriptor of disease states \cite{scheffer2024}. Central to this framework is the amygdala, which occupies a core position within a broader corticolimbic and autonomic network \cite{tovote2015,janak2015}. Sensory information reaches the lateral amygdala via subcortical and corticothalamic pathways, propagates through the basal and intercalated nuclei and converges on central outputs that project to hypothalamic and brainstem centres controlling heart rate, blood pressure, and other autonomic responses \cite{khalil2023,schaeuble2022}. In healthy individuals the system supports graded, transient and homeostatically regulated threat responses, whereas in chronic anxiety and depression identical stimuli evoke amplified amygdala responses, weakened prefrontal control, and delayed, often incomplete recovery \cite{vaccarino2024,cheng2022,berboth2021,etkin2015}. Understanding the mechanisms by which a single circuit governs both adaptive vigilance and maladaptive hypereactivity remains fundamental to developing mechanism-based interventions, which rely on parallel advances in sensing and modeling.
 Digital‑phenotyping and personal‑sensing studies show that behavioral, physiological, and self‑report data from smartphones and wearables yield continuous indicators of stress, anxiety, and subthreshold mood disturbance \cite{choi2024,booth2022,zhao2025,torous2021,insel2023}. Simultaneously, digital‑twin and mechanistic‑modeling frameworks in cardiology and oncology map parameter shifts to measurable physiological trajectories for clinical decision support \cite{bjornsson2019,laubenbacher2022}. The critical research gap at this intersection is the lack of a mechanistic model of central circuit that generates the autonomic profiles being monitored.  Without such model, an elevation in heart rate or a change in blood‑pressure patterns at the wrist cannot be traced back to distinct variations in prefrontal regulation, perceived stress load, or baroreflex function.\\
 Bridging this multiscale gap presents a formidable challenge because the available tools operate at different levels of abstraction. At the cellular level, conductance‑based neuron models in the Hodgkin–Huxley framework relate synaptic currents, ion‑channel kinetics, and membrane capacitance to voltage trajectories and spiking dynamics \cite{hodgkin1949,hausser2000}. At the cognitive and clinical level, self‑report instruments such as the Connor–Davidson Resilience Scale (CD-RISC) and the Perceived Stress Scale-10 (PSS-10) quantify resilience and perceived stress \cite{connor2003,cohen1983,liu2020}. While these metrics reliably  distinguish clinical populations, and correlate with cardiovascular stress responsivity \cite{wojujutari2024,vaccarino2024}, existing computational models remain confined to a single side of this gap. System‑level models treat the amygdala and its downstream pathways as an electrophysiological black box driven by abstract inputs \cite{matsubara2021,matsubara2019}. Meanwhile, high‑fidelity biophysical models collapse its contribution into a rigid fixed‑threshold mechanism \cite{lindsey2009,cline2004}. This latter neglects slow, history‑dependent sensitization, through which cumulative stress reshapes cellular excitability \cite{vaccarino2024,berboth2021}. Consequently, existing model exposes no parameters matching established clinical constructs and lacks the coupled cardiovascular outputs needed for validation against wearable recordings.  Capturing both chronic stress-amplified threat appraisal and its downstream propagation to cardiovascular dynamics within a single framework remains a significant challenge.  We address this gap by developing and validating a mechanistic model of the amygdala–hypothalamus–cardiovascular pathway structured as an integrated bio‑signal modeling framework. A single amygdala principal neuron is modeled as a conductance‑based unit in which gating variable is reinterpreted as a slow, history‑dependent internal state integrating prior activation over timescales relevant to stress‑induced sensitization \cite{hodgkin1949,hausser2000}. Three modulators, coping capacity, perceived stress load, and prefrontal regulatory strength, enter the synaptic and threshold dynamics directly, establishing a transparent mapping from clinically interpretable quantities to model parameters. Threat acknowledgment is quantified via a graded sigmoid function of the difference between membrane potential and a modulator‑dependent adaptive threshold. This suprathreshold component drives a hypothalamic integrator with baroreflex nucleus tractus solitarius (NTS) feedback. The resulting architecture yields time‑varying heart‑rate and blood‑pressure trajectories that can be checked against published stress‑response ranges and independent datasets \cite{hosseini2022,novak2010}.\\
The proposed framework advances the field through four principal contributions.
\begin{enumerate}

\item \textbf{Mechanistic Amygdala Model with Slow Adaptation and adaptive threshold.} We adapt the Hodgkin–Huxley conductance-based framework to represent an amygdala neuron \cite{hodgkin1949,hausser2000}. This approach preserves biophysical interpretability through a reinterpreted gating variable. The gating variable serves as a compact substrate for slow, history-dependent adaptation. Simultaneously an adaptive, state-dependent threshold regulates graded threat acknowledgment \cite{berboth2021}. This mechanism captures the transition from rapid adaptation to sustained sensitization that distinguishes healthy regimes from those of anxiety and depression \cite{vaccarino2024}.
    
\item \textbf{Psychometric to Parameter Mapping of Regulatory Regimes.} We map coping capacity, perceived stress load, and prefrontal regulatory strength directly onto the model parameters. These modulators are assessed from clinical scales (CD-RISC, PSS-10) and a neuroimaging-derived prefrontal index \cite{berboth2021,etkin2015,connor2003,cohen1983,liu2020,wojujutari2024,matsubara2021,matsubara2019}. Thus the model is driven by interpretable clinical dimensions rather than abstract parameters, yielding three regimes (strong, moderate, weak) for healthy, subclinical, and pathological phenotypes along a continuous spectrum.

\item \textbf{Baroreflex‑coupled hypothalamic integrator producing autonomic outputs.} We couple the graded amygdala output to a hypothalamic integrator and a baroreflex feedback loop \cite{schaeuble2022,catrambone2023,sadoun2025,blasi2005,hyndman1996}. This integration generates heart‑rate and blood‑pressure trajectories consistent with acute sympathetic activation and subsequent recovery, including regime‑dependent peak excursions and baseline return. These outputs provide a direct interface for wearable sensors and for comparison with independent datasets spanning mental‑health monitoring, occupational stress, and functional rehabilitation.

\item \textbf{Cross-Scale Validation in Sparse-Sensor Environments.} To address the scarcity of simultaneous  human cellular, psychometric, and continuous autonomic data, we implement a convergent-validation strategy. This framework bridges biological scales by coupling internal sensitivity analyses (parameter perturbations and component ablations) with external benchmarking against three independent datasets. Consequently, this approach decouples the model's structural strengths from calibration artifacts, verifying its ability to reproduces real-world autonomic stress dynamics.

\end{enumerate}

Integrated into a single framework, these elements deliver a tractable set of equations where elevated reactivity, reduced discriminability, and impaired homeostatic return emerge from defined, clinically interpretable modulator variations. This architecture enables central threat acknowledgment and peripheral autonomic profiles to be jointly modeled as signals within a closed‑loop system. The paper proceeds as follows. Section~II presents the model formulation and  parameter mapping. Section~III details the stimulation protocol, simulation setup, and regulatory regimes. Section~IV evaluates the amygdala, hypothalamic, and autonomic results alongside their physiological interpretation. Section~V provides robustness analyses, ablations, external empirical validation, and translational scope of the work. Section~VI concludes with the primary findings and future research directions.

\section{MODEL}\label{sec:model}

\subsection{Sensory Drive and Amygdala Conductance‑Based Neuron}\label{sec:sensorydrive}
\begin{figure}[htbp]
  \centering
  \includegraphics[width=\linewidth]{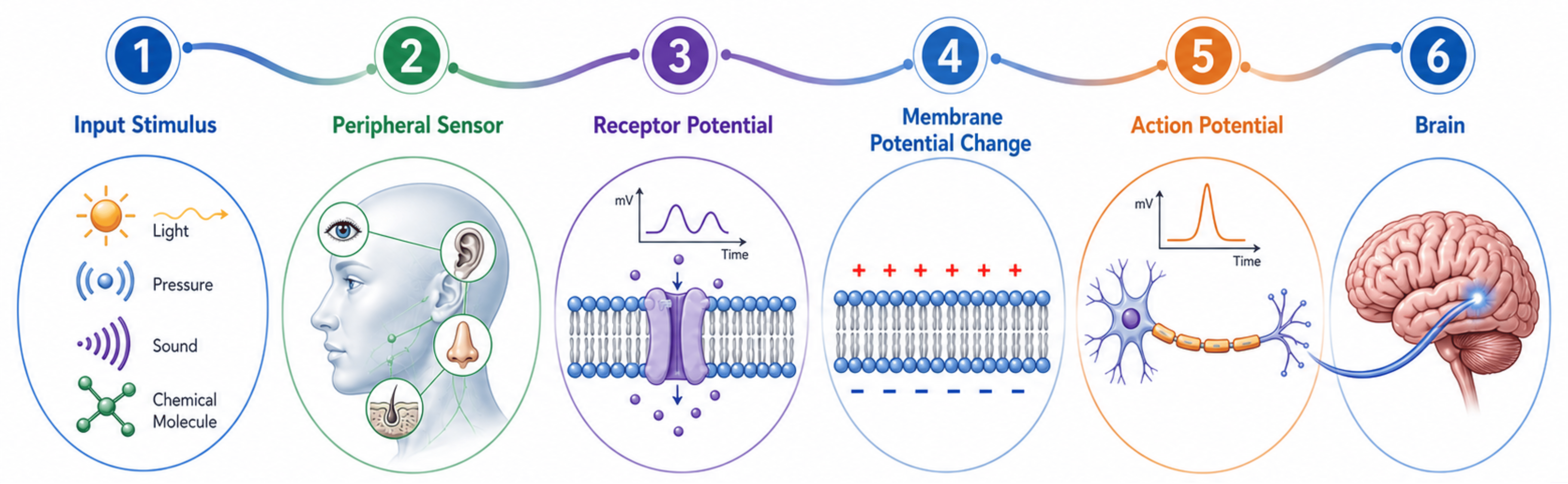}
   \caption{Information flow from external environmental stimuli to central neural processing}
  \label{fig:infomation}
\end{figure}

External stimuli across different modalities (e.g., light, pressure, sound, chemical concentration) are first mapped to a common graded receptor potential $R(t)\!\in\![0,100]$~mV via stimulus‑gated ion channels. As the rise and decay time of $R(t)$ vary with stimulus intensity and duration, the potential encodes both the magnitude and the persistence of the input. Rather than modeling modality‑specific transduction pathways in detail, we treat $R(t)$ as the effective sensory drive to upstream neurons, which is approximately proportional to their subthreshold membrane potential (Fig~\ref{fig:infomation}).

When this sensory drive depolarizes the membrane beyond a threshold near $−55$ mV, action potentials are generated and conveyed to the amygdala via two functionally distinct routes: a fast subcortical pathway through the lateral thalamus supporting rapid pre‑conscious reactions, and a slower corticothalamic pathway that permits deliberate interpretation and prefrontal inhibition. Both pathways deliver glutamatergic excitation to the amygdala network. We model an amygdala principal neuron as a single isopotential compartment with membrane capacitance $C_m$, receiving a passive leak conductance, an inhibitory synaptic conductance, and a modulatory excitatory synaptic conductance. Within this conductance‑based framework, the membrane voltage does not follow the raw stimulus directly.Instead, it emerges from the balance of ionic and synaptic currents flowing across the membrane as a consequence of presynaptic activity along converging sensory pathways. This fundamental electrical behavior of the isopotential compartment is governed by the current-balance equation derived from Kirchhoff's current law:
\begin{equation}
  C_m \frac{dV_m(t)}{dt} 
  = I_{\mathrm{L}}(t)
  + I_{\mathrm{syn,inhib}}(t)
  + I_{\mathrm{syn,exc}}(t)
\label{eq:current_balance}
\end{equation}
where each current $I = g \cdot (V_m - E)$ is expressed as the product of conductance $g$ (time or voltage dependent) and corresponding driving force $E$ (reverse potential). We set the specific membrane capacitance to $C_m = 1\,\mu\text{F/cm}^2$ throughout, adhering to the canonical Hodgkin--Huxley baseline~\cite{hodgkin1949,hausser2000}, such that all currents represent current densities expressed in $\mu\text{A/cm}^2$. The passive leak current, $I_{\mathrm{L}}(t)$, accounts for steady, voltage-independent ion flux through non-gated channels. Because potassium permeability dominates the resting state, the leak reverse potential is set to $E_{\mathrm{L}} = -70$~mV, yielding:
\begin{equation}
  I_{\mathrm{L}}(t) = -g_{\mathrm{L}} \bigl[V_m(t) - E_{\mathrm{L}}\bigr]
  \label{eq:leak}
\end{equation}

where $g_L$  a constant leak conductance and the sign convention signifies that outward currents are negative. The leak conductance directly establishes the passive membrane time constant, $\tau_m = C_m/g_L$, which determines the relaxation rate of $V_m(t)$ toward $E_L$ upon input removal. Together with synaptic inputs, $\tau_m$ determines the overall gain and system stability of the neuron. Hence, a larger $g_{\mathrm{L}}$ creates a stiffer system characterized by attenuated voltage excursions, whereas a lower $g_{\mathrm{L}}$ allows wider dynamic trajectories around the resting potential. Inhibitory synaptic currents in the amygdala arise primarily from local GABAergic interneurons and intercalated cells, with additional regulatory contributions from glycinergic and prefrontal inputs \cite{janak2015,tovote2015}. The Opening of Cl$^-$/K$^+$-permeable channels hyperpolarises the membrane and shunts incoming excitation. We model the net inhibitory current as

\begin{equation}
  I_{\mathrm{syn,inhib}}(t)
  = -\,g_{\mathrm{in}}\,h(t)\,\bigl[V_m(t) - E_{\mathrm{syn}}\bigr]
  \label{eq:inhib}
\end{equation}

where  $E_{\mathrm{syn}}\!\approx\!-70$~mV , $g_{\mathrm{in}}h(t)$ is the maximum baseline inhibitory conductance and the dimensionless factor $h(t)$ encodes history‑dependent modulation of inhibition. The slow internal state variable $h(t)$ acts as a multiplicative factor that adjusts effective inhibitory strength based on prior activation and is analyzed in detail in next subsection. The excitatory drive is dominated by glutamatergic input from thalamus and sensory cortex. We model this input as a sensory current  $I_{\mathrm{sensory}}(t)$ derived from the receptor potential and gated by prefrontal regulatory strength $R_{\mathrm{PFC}}(t)$:
\begin{align}
  I_{\mathrm{syn,exc}}(t) &= I_{\mathrm{sensory}}(t)\,I_{\mathrm{PFC}}(t), \label{eq:exc} \\
  \, I_{\mathrm{PFC}}(t) &= 1 - R_{\mathrm{PFC}}(t),  I_{\mathrm{sensory}}(t) = K_s R(t) \nonumber
\end{align}

where  $K_s$ is a gain parameter setting the scale of sensory input and $R_{\mathrm{PFC}}(t)\!\in\![0,1]$ represents the fraction of excitatory drive suppressed by prefrontal control. High $R_{\mathrm{PFC}}$ values strongly attenuates excitatory input to the amygdala, whereas low $R_{\mathrm{PFC}}$ values allows sensory drive to pass largely unfiltered.\\
Substituting \eqref{eq:leak}-\eqref{eq:exc} into \eqref{eq:current_balance} yields the modulated membrane equation
\begin{equation}
\begin{split}
  C_m\,\frac{dV_m(t)}{dt} = 
  &-g_L\bigl[V_m - E_L\bigr] - g_{\mathrm{in}}h(t)\bigl[V_m - E_{\mathrm{syn}}\bigr] \\
  &+ I_{\mathrm{sensory}}(t)\bigl[1 - R_{\mathrm{PFC}}(t)\bigr]
\end{split}
\label{eq:full}
\end{equation}
which we integrate numerically using a forward‑Euler scheme with time step $\Delta t$:
\begin{equation}
\begin{split}
  V_i = V_{i-1} + \frac{\Delta t}{C_m}\,I_{\mathrm{total}}(i),\\
  I_{\mathrm{total}}
  = I_{\mathrm{ion}} + I_{\mathrm{syn,inhib}} + I_{\mathrm{syn,exc}}.
  \end{split}
  \label{eq:euler}
\end{equation}

In this formulation, the amygdala neuron acts as a context-dependent integrator that combines modality-independent sensory drive $R(t)$, history-dependent inhibition $h(t)$, and top-down prefrontal regulation $R_{\text{PFC}}$ into a continuously evolving membrane potential $V_{\text{m}}(t)$. This voltage serves as the internal decision variable for the subsequent thresholding, threat-acknowledgment, and autonomic integration stages described in the following subsections.

\subsection{Internal State and Cognitive‑Affective Modulation}\label{sec:internalstate}
Classical Hodgkin–Huxley model describe inactivation of voltage‑gated sodium channels by a gating variable $h(t)$ that obeys first‑order voltage‑dependent kinetics \cite{hodgkin1949,hausser2000}. In our framework, we retain this mathematical structure but reinterpret $h(t)$ as a slow internal state that captures the history‑dependent modulation of amygdala responsiveness on timescales relevant for sensitization, adaptation, and stress‑induced plasticity. Rather than representing the fraction of available sodium channels, $h(t)$  encodes the cumulative effect of prior activation on the cell’s current sensitivity. The amygdala membrane dynamics are therefore determined jointly by the instantaneous sensory drive $I_{sensory}(t)$ and this internal state $h(t)$, consistent with clinical evidence that repeated stress-related stimulation progressively modifies amygdala reactivity in anxiety and depression. Accordingly, the temporal evolution of $h(t)$ follows first-order relaxation toward a voltage-dependent steady state $h_\infty(V_m)$:
\begin{equation}
\begin{split}
  \frac{dh(t)}{dt} &= \frac{h_\infty(V_m) - h(t)}{\tau_h(V_m)}, \\[1ex]
  \ h_\infty(V_m) &= \frac{1}{1 + \exp\!\big[-(V_m - V_{mo})/k\bigr]}
\end{split}
\label{eq:h_dynamics}
\end{equation}
where $V_{m0} = -40\text{ mV}$ is the half-activation voltage, and $k \in [5, 15]\text{ mV}$ represents the slope factor ( smaller $k$ values yield steeper, switch-like dependence while, larger $k$ values yield a more graded response). The voltage-dependent time constant $\tau_h(V_m)$ determines the history timescale: values on the order of 20–40~ms correspond to rapid adaptation, values around 80–120~ms corresponds to behaviorally relevant integration of recent activity, and values $\ge 200\text{ ms}$ model longer-term sensitization. Because $V_m(t)$ evolves on a fast millisecond scale while $h(t)$ relaxes over tens to hundreds of milliseconds, the model exhibits a slow-fast separation where $h(t)$ tracks the cumulative voltage history. This structural feature underpins the history-dependent modulation utilized in our framework. In addition to this internal state, we introduce three cognitive-affective modulators that represent state-like properties of the individual: coping capacity $C(t)$, perceived stress load $P(t)$, and prefrontal regulatory strength $R_{PFC}(t)$. These quantities are derived from clinical psychometric and neuroimaging measures that vary slowly compared with the millisecond-scale dynamics of the simulated amygdala. Consequently, within each local simulation episode, we treat $C(t)$, $P(t)$, and $R_{PFC}(t)$ as quasi-static parameters that characterize the current individual rather than fast dynamic variables. The time dependence notation $(t)$ emphasizes that these variables may evolve over much longer timescales across episodes (e.g., due to chronic stress progression or clinical recovery), though such macro-evolution is not explicitly modeled within this specific framework.

Coping capacity $C(t)$ quantifies the regulatory resources available for evaluating and recovering from stressors and is operationalized via the Connor--Davidson Resilience Scale (CD-RISC, summed and normalized to a bounded range) \cite{connor2003,wojujutari2024}. Higher $C(t)$ strengthens inhibitory control by increasing inhibitory conductance, reducing depolarization, elevating thresholds, and damping sensitization via $h(t)$. Conversely, lower $C(t)$ results in weaker regulation and disproportionately strong reactions to weak stimuli. Perceived stress load $P(t)$ reflects the subjective sense that ongoing demands exceed available resources and is operationalized by the Perceived Stress Scale-10 (PSS-10) \cite{cohen1983,liu2020}. Higher $P(t)$ weakens inhibition, depolarizes the membrane, lowers the effective threshold, and increases the likelihood for threat acknowledgment. It further attenuates baroreflex feedback by reducing baroreceptor and nucleus tractus solitarius sensitivity. Prefrontal regulatory strength $R_{\text{PFC}}(t)\in [0,1]$ represents top-down control by medial and dorsolateral prefrontal cortex. Values approaching unity signifies the suppressing of excitatory drive to the amygdala, where as lower values reflect the attenuated prefrontal-amygdala coupling observed in chronic anxiety and depression.

\begin{figure}[htbp]
  \centering
  \includegraphics[width=\linewidth]{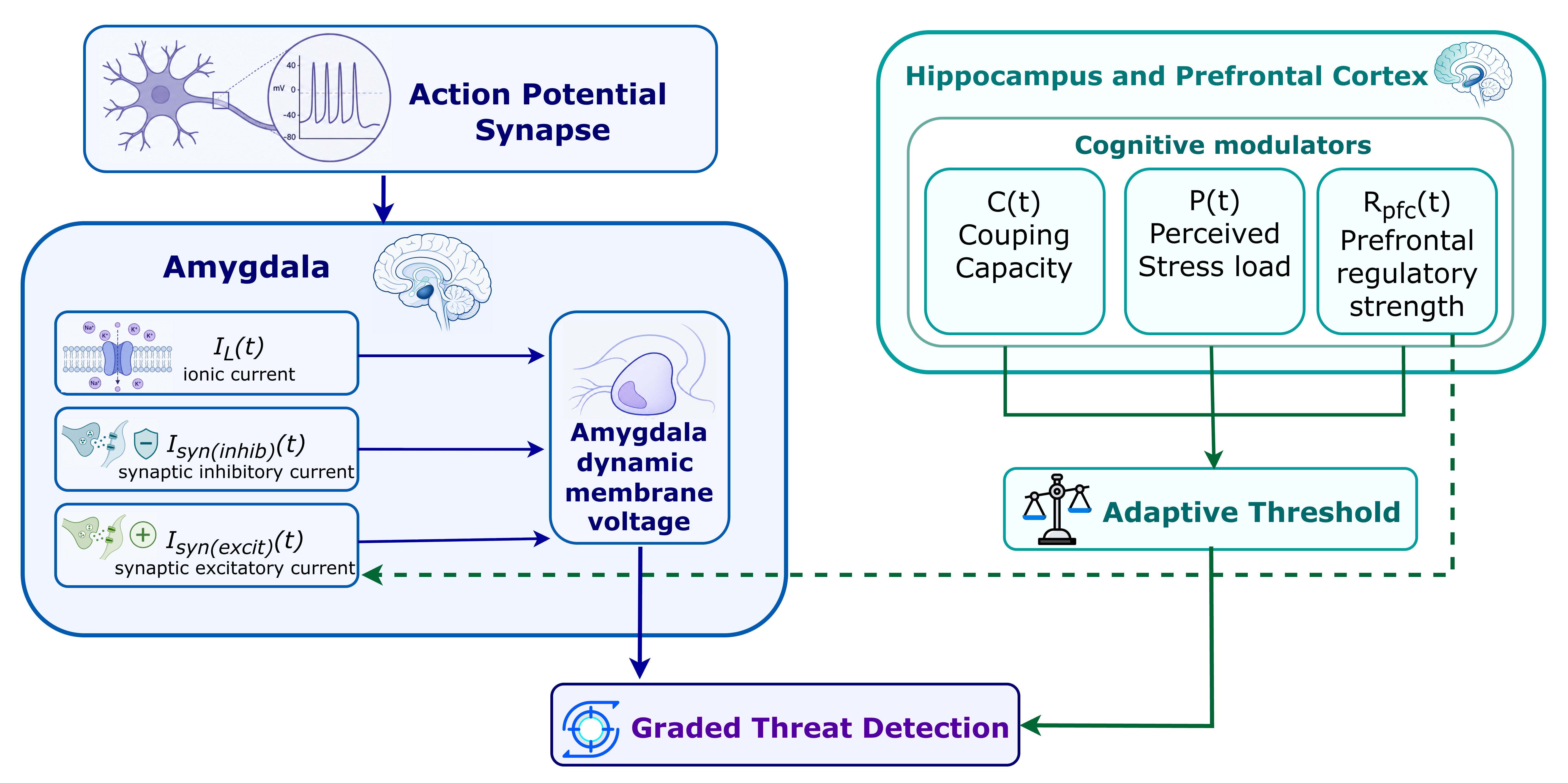}
   \caption{Computational schematic of amygdala threat processing with cognitive modulation.}
  \label{fig:amygdala threat process}
\end{figure}

These modulators enter the dynamics by scaling the synaptic currents (Section~\ref{sec:sensorydrive}) and shifting excitability. Coping capacity $C(t)$ enhances inhibitory tone by increasing the effective inhibitory conductance $g_{\text{inh}}(t)$, whereas perceived stress $P(t)$ reduces inhibitory control and amplifies the incoming sensory currents. Prefrontal regulatory strength $R_{\text{PFC}}(t)$ gates the excitatory current through the factor $I_{\text{PFC}}(t) = 1 - R_{\text{PFC}}(t)$, where strong prefrontal regulation attenuates amygdala drive while weak regulation allows sensory input to pass unfiltered. Together, $h(t)$, $C(t)$, $P(t)$, and $R_{\text{PFC}}(t)$ set the effective gain and the excitation-inhibition balance at the amygdala neuron (Fig~\ref{fig:amygdala threat process}). In addition these modulators shape the adaptive threshold and hypothalamic-autonomic pathway to produce context-dependent threat acknowledgment and cardiovascular responses across regimes.

\subsection{Threat Acknowledgement and Autonomic Output}\label{sec:threatacknow}
\begin{figure}[htbp]
  \centering
  \includegraphics[width=\linewidth]{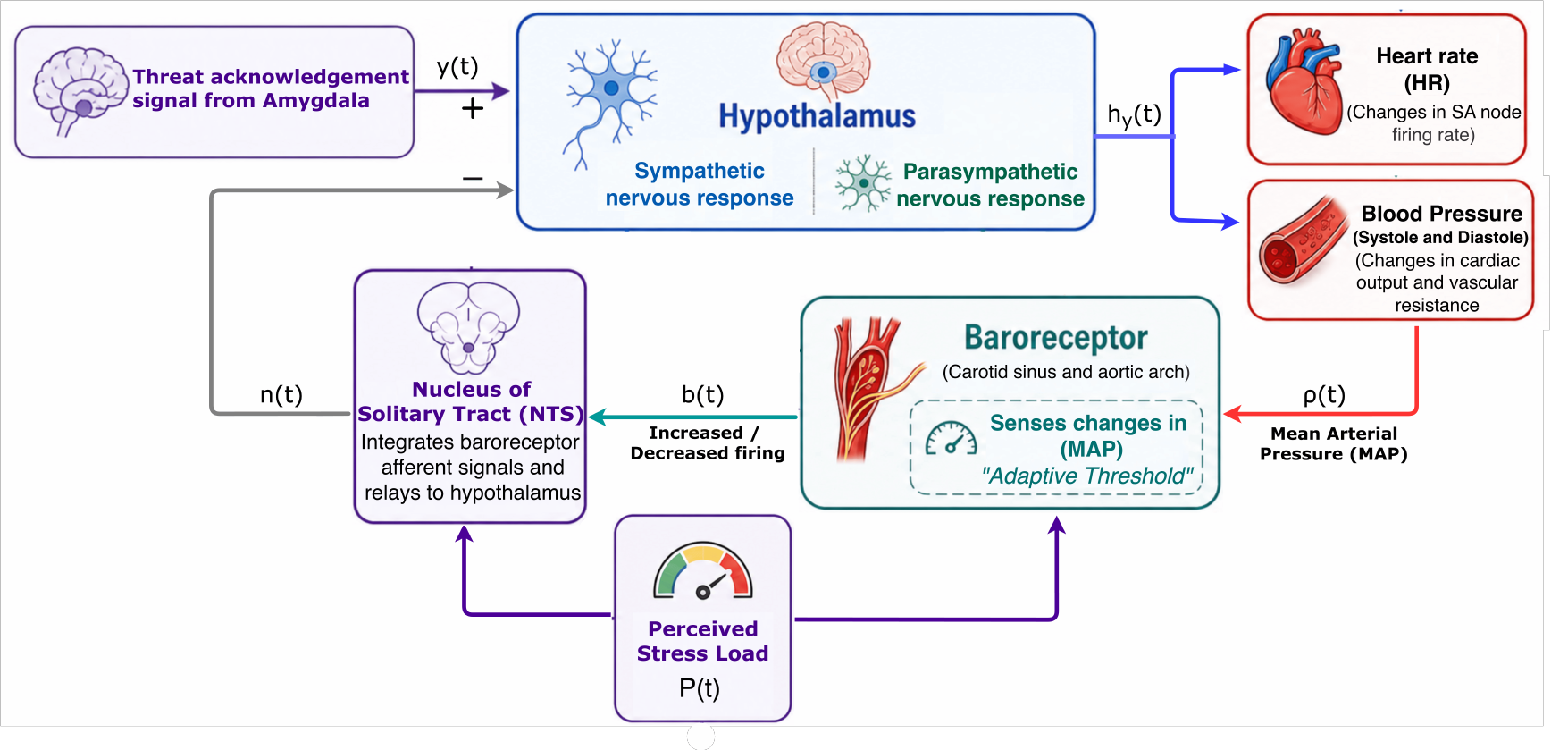}
   \caption{Closed‑loop baroreflex feedback linking hypothalamic regulation, cardiovascular outputs, and stress‑dependent modulation.}
  \label{fig:hypothalmus feedback}
\end{figure}

For an integrative structure such as the amygdala, a fixed firing threshold is physiologically inadequate because excitability depends on prior activity, neuromodulation, and stress‑induced changes in membrane properties. We therefore introduce an adaptive threshold $\theta_0$ that depends linearly on the cognitive‑affective modulators defined in Section~\ref{sec:internalstate},
\begin{equation}
  \theta(t)
  = \theta_0 + K_c C(t) - K_p P(t) + K_r R_{\mathrm{PFC}}(t),
  \label{eq:threshold}
\end{equation}
where $\theta_0 = -55$~mV represents the baseline threshold. Stronger coping capacity and prefrontal regulatory strength elevates $\theta_(t)$ , whereas higher perceived stress lowers it. Stratifying each modulator into normal, moderate, and severe regimes yields corresponding threshold levels $\theta_{\text{nor}}$, $\theta_{\text{mod}}$, and $\theta_{\text{sev}}$, to represent distinct anxiety and depression profiles.
The amygdala output is defined as a graded threat-acknowledgment signal obtained by passing the difference of membrane potential $V_m(t)$ and adaptive threshold $\theta(t)$ through a sigmoidal nonlinearity:
\begin{equation}
  A(u, t) = \sigma\!\bigl(u\,\bigl[V_m(t)\!-\!\theta(t)\bigr]\bigr),
  \quad \sigma(x) = (1+e^{-x})^{-1},
  \label{eq:threat}
\end{equation}
with gain $u > 0$ calibrated from data. This mapping is consistent with standard signal-detection formulations which converts a noisy decision variable into a response probability: $A(u,t) \to 0$ when $V_m(t) \ll \theta(t)$, $A(u,t) \to 1$ when $V_m(t) \gg \theta(t)$, and $A(u,t) \approx 0.5$ indicates borderline threat acknowledgment. Rather than an all-or-none decision, this signal is relayed from the central nucleus of the amygdala to the hypothalamus, which integrates amygdala drive with cardiovascular feedback to generate autonomic output. The suprathreshold input to the hypothalamus is defined as:
\begin{equation}
  y(t) =
  \begin{cases}
    0,         & A(u, t) < 0.5 \\
    A(u, t),   & A(u, t) \geq 0.5
  \end{cases}
  \label{eq:y_pwr}
\end{equation}
so that only threat acknowledgments above a behaviorally meaningful midline contribute to autonomic recruitment. The value $0.5$ is a modeling convenience rather than a physiological constant. The hypothalamus is modeled as a central autonomic integrator whose state $h_{\text{y}}(t)$ evolves under the combined influence of amygdala drive $y(t)$, baroreflex-mediated inhibitory feedback $n(t)$, and a passive decay:
\begin{equation}
  \frac{dh_y(t)}{dt}
  = \alpha\,y(t)\,\bigl[1 - h_y(t)\bigr]
  - \beta\,n(t)\,h_y(t)
  - \gamma\,h_y(t).
  \label{eq:hypo_ode}
\end{equation}
The first term saturates excitatory influence as $h_{\text{y}}(t) \to 1$, representing the ceiling of sympathetic output. The second implements baroreflex feedback that suppresses further recruitment and the third is passive decay toward baseline. The gains $\alpha$, $\beta$, and $\gamma$ are modulated by the cognitive-affective variables to reflect chronic stress and modified regulation.
Baroreflex feedback is generated via an explicit baroreceptor-nucleus tractus solitarius (NTS) pathway. The baroreceptor output $b(t)$ is modeled as a sigmoidal function of mean arterial pressure $\rho(t)$ relative to an adaptive reference point $\phi(t)$,
\begin{equation}
b(t) = \sigma \left( \frac{\rho(t) - \phi(t)}{\eta_b} \right), \quad \rho = \frac{2\text({DBP}) + \text{SBP}}{3}
\end{equation}
where $\eta_b$ is a sensitivity parameter. Both the reference point and sensitivity shift under perceived stress load $P(t)$,
\begin{equation}
\phi(t) = \phi_0 - K_b.P(t), \quad \eta_b = \eta_{b0} - K_{bs}. P(t)
\end{equation}
with coupling constants $K_b$ and $K_{bs}$ governing these stress-induced changes. 
Increased perceived stress thus elevates the effective reference pressure and reduces sensitivity, weakening baroreceptor responses. The baroreceptor signal is relayed to the NTS, modelled as a leaky integrator,
\begin{equation}
\frac{dn(t)}{dt} = K_n .b(t) - \delta n(t).
\end{equation}
with gain $K_n$ and decay rate $\delta$. Under forward-Euler discretisation 
with time step $\Delta t$, the NTS state updates as
\begin{equation}
n(t) = n(t - \Delta t) + \Delta t [K_n .b(t) - \delta n(t - \Delta t)]
\end{equation}
and the resulting $n(t)$ constitutes the inhibitory feedback term in~\ref{eq:hypo_disc} (Fig~\ref{fig:hypothalmus feedback}). 

Sustained high perceived stress, reduced coping capacity, and impaired 
prefrontal control are represented as reductions in $\beta$ and $\gamma$. 
The increase in $\alpha$, describes an amplified amygdala drive and weakened peripheral negative 
feedback in chronic stress.
We discretise the hypothalamic dynamics (11) using a forward-Euler update with a time step $\Delta t = 0.1$~ms:
\begin{equation}
\begin{split}
  h_y(t) &
  = h_y(t - \Delta t)
  + \Delta t\,\Bigl[
     \alpha\,y(t)\,\bigl(1 - h_y(t - \Delta t)\bigr)\\
   & - \beta\,n(t)\,h_y(t - \Delta t)
   - \gamma\,h_y(t - \Delta t)
  \Bigr]
  \end{split}
  \label{eq:hypo_disc}
\end{equation}
Changes in the hypothalamic state are then mapped to observable cardiovascular responses, heart rate (HR), systolic blood pressure (SBP), and diastolic blood pressure (DBP), as deviations from individual baseline values:

\begin{equation}
\begin{split}
HR(t) &= HR_{\text{rest}} + \Delta HR(t),\\
&\Delta HR(t) = g_{hr} \cdot (h_y(t) - h_{y0}) \\
SBP(t) &= SBP_{\text{rest}} + \Delta SBP(t),\\ 
&\Delta SBP(t) = g_{sbp} \cdot (h_y(t) - h_{y0}) \\
DBP(t) &= DBP_{\text{rest}} + \Delta DBP(t), \\
&\quad  \Delta DBP(t) = 0.6 \cdot \Delta SBP(t)\\
\end{split}
\label{eq:hypo_out}
\end{equation}

where $\text{HR}_{\text{rest}}$, $\text{SBP}_{\text{rest}}$, and $\text{DBP}_{\text{rest}}$ denote resting values and $g_{\text{hr}}$, $g_{\text{sbp}}$ are gain factors mapping hypothalamic activation to haemodynamic changes. Through the combined action of the adaptive threshold $\theta(t)$, graded threat acknowledgement 
$A(u,t)$, hypothalamic integration, and baroreflex modulation, the model 
links amygdala membrane dynamics and cognitive-affective modulators to 
individual-specific patterns of autonomic response.

\section{Simulation Protocol} \label{sec:simulationprtocol}
\subsection{Fixed Parameters and Regulatory Regimes} \label{sec:fixedpara}
All simulations use the integrated model described in Section~\ref{sec:model} with parameters listed in table~\ref{tab:params} and three regulatory regimes (strong, moderate, weak) defined in table~\ref{tab:regimes}. The baseline biophysical parameters follow canonical conductance-based neuron values, while the threshold scaling coefficients and readout gain ensure the amygdala output operates within the graded portion of the sigmoid transfer function.

\begin{table}[htbp]
  \caption{Fixed biophysical and model parameters following standard Hodgkin–Huxley framework.}
  \label{tab:params}
  \centering
  \footnotesize 
  \setlength{\tabcolsep}{3pt} 
  \begin{tabular}{lp{3.15cm}rl}
    \toprule
    \textbf{Symbol} & \textbf{Description} & \textbf{Value} & \textbf{Units} \\
    \midrule
    $C_m$            & Membrane capacitance               & $1.00$  & $\mu\mathrm{F\,cm^{-2}}$ \\
    $g_L$            & Leak conductance                   & $0.10$  & $\mathrm{mS\,cm^{-2}}$ \\
    $E_L$            & Leak reversal potential            & $-70$   & mV \\
    $g_{\mathrm{in}}$& Peak-conductance-inhibt       & $0.30$  & $\mathrm{mS\,cm^{-2}}$ \\
    $E_{\mathrm{syn}}$ & Inhibitory reversal potential    & $-70$   & mV \\
    $V_{m0}$         & Half-max voltage of $h_\infty$     & $-40$   & mV \\
    $k$              & Slope factor of $h_\infty$         & $8$     & mV \\
    $K_s$            & Sensory current scaling            & $0.6$   & $\mathrm{mS\,cm^{-2}}$  \\
    $\theta_0$       & Baseline threshold                 & $-55$   & mV \\
    $K_c, K_p, K_r$  & Threshold scaling coeffs.          & $4, 5, 4$ & mV \\
    $u$              & Sigmoidal readout gain             & $0.20$  & $\mathrm{mV^{-1}}$ \\
    $K_b, K_{\mathrm{bs}}, K_n$  & Baroreflex/NTS coeffs. & \hspace{-1em}$0.30, 0.075, 0.045$ & \\
    \bottomrule
  \end{tabular}
\end{table}
The three regulatory regimes represent strong (healthy), moderate (sub-clinical), and weak (anxiety/depression) control. These regimes are implemented by varying the cognitive-affective modulators $(R_{\text{PFC}}, C, P)$, the adaptation time constant $\tau_{\text{h}}$, and the hypothalamic/baroreflex coefficients $\alpha, \beta, \gamma, \delta$.

\begin{table}[htbp]
  \caption{Cognitive‑affective modulator and hypothalamic/baroreflex parameter settings for the varied regulatory regimes~\eqref{eq:hypo_ode} }
  \label{tab:regimes}
  \centering
  \footnotesize 
  \setlength{\tabcolsep}{2.5pt} 
  \begin{tabularx}{\columnwidth}{>{\raggedright\arraybackslash}Xcccccccc}
    \toprule
    \textbf{Regime} & $R_{\mathrm{PFC}}$ & $C(t)$ & $P(t)$ & $\tau_h$ & $\alpha$ & $\beta$ & $\gamma$ & $\delta$ \\
    \midrule
    Strong (healthy)          & $0.75$ & $0.75$ & $0.25$ & $40$  & $0.35$ & $0.23$ & $0.07$ & $0.17$ \\
    Moderate (sub-clin.)      & $0.50$ & $0.50$ & $0.50$ & $100$ & $0.60$ & $0.14$ & $0.04$ & $0.09$ \\
    Weak (anx./depr.)         & $0.25$ & $0.25$ & $0.75$ & $220$ & $0.98$ & $0.06$ & $0.02$ & $0.04$ \\
    \bottomrule
  \end{tabularx}
\end{table}

 As detailed in Section~\ref{sec:threatacknow}, the weak control regime decreases $\beta$ and $\gamma$, while increasing $\alpha$, to model stress induced baroreflex degradation and prolonged hypothalamic activation.

\subsection{Numerical Integration and Stimulus Design}  \label{sec:Num Integ & Stimulus}
The coupled neuron–hypothalamus system is integrated using a forward-Euler scheme with time step $\Delta t = 0.1$~ms over a total simulation window $T = 5000$~ms. Initial conditions are set to $V_0 = -55$~mV, $h_0 = h_{\infty}(V_0)$, $h_y(0) = 0.01$, and $n(0) = 0.04$, representing a baseline sympathetic tone. The external stimulus consists of five 400~ms rectangular pulses with an amplitude 0.5, applied at 0.8, 1.5, 2.2, 2.9, and 3.6~s, superimposed on an Ornstein–Uhlenbeck noise process($\tau = 80$~ms, $\sigma = 0.12$, clipped to $[0,1]$). The receptor potential $R(t) $ is generated with rise and decay time constants $\tau_r = 15$~ms and $\tau_d = 50$~ms and a 40~mV peak. The same noise realization is reused across all regimes to isolate the effects of the modulators and regime-specific parameters. To embed the roles of coping capacity $C$ and perceived stress $P$ in inhibitory tone, we define an effective inhibitory conductance as: 

\begin{equation}
g_{\mathrm{in,eff}} = g_{\mathrm{in}}(1 + 0.4C - 0.5P)
\end{equation}
The adaptation time constant $\tau_h$ is set to 40, 100, and 220~ms in the strong, moderate, and weak regimes respectively (Table~\ref{tab:regimes}), controlling the cross-pulse accumulation in the slow internal state.

Parameters are separated into physiology baselines and calibration coefficients. Physiological parameters ($C_m$, $g_L$, $E_L$, $E_{\text{syn}}$, $V_{m0}$, resting cardiovascular values, and input-noise time constant) follows standard conventions. Calibration coefficients ($K_c, K_p, K_r, K_b, K_{\text{bs}}, K_n$ and readout gain $u$) are tuned maintain the amygdala readout within the graded sigmoidal regime. Accordingly, the ordering, monotonicity, and structural roles of modulators and regimes are robust, whereas the absolute magnitudes of $A(t)$, its dispersion $\sigma_A$, and variance ratios are calibration-dependent, as qualified by the robustness and sensitivity analyses below.

\section{Results and Discussion}\label{sec:result}

\subsection{Overview of Regime Behavior} \label{sec:overview regime}

\begin{figure}[htbp]
  \centering
  \includegraphics[width=\linewidth]{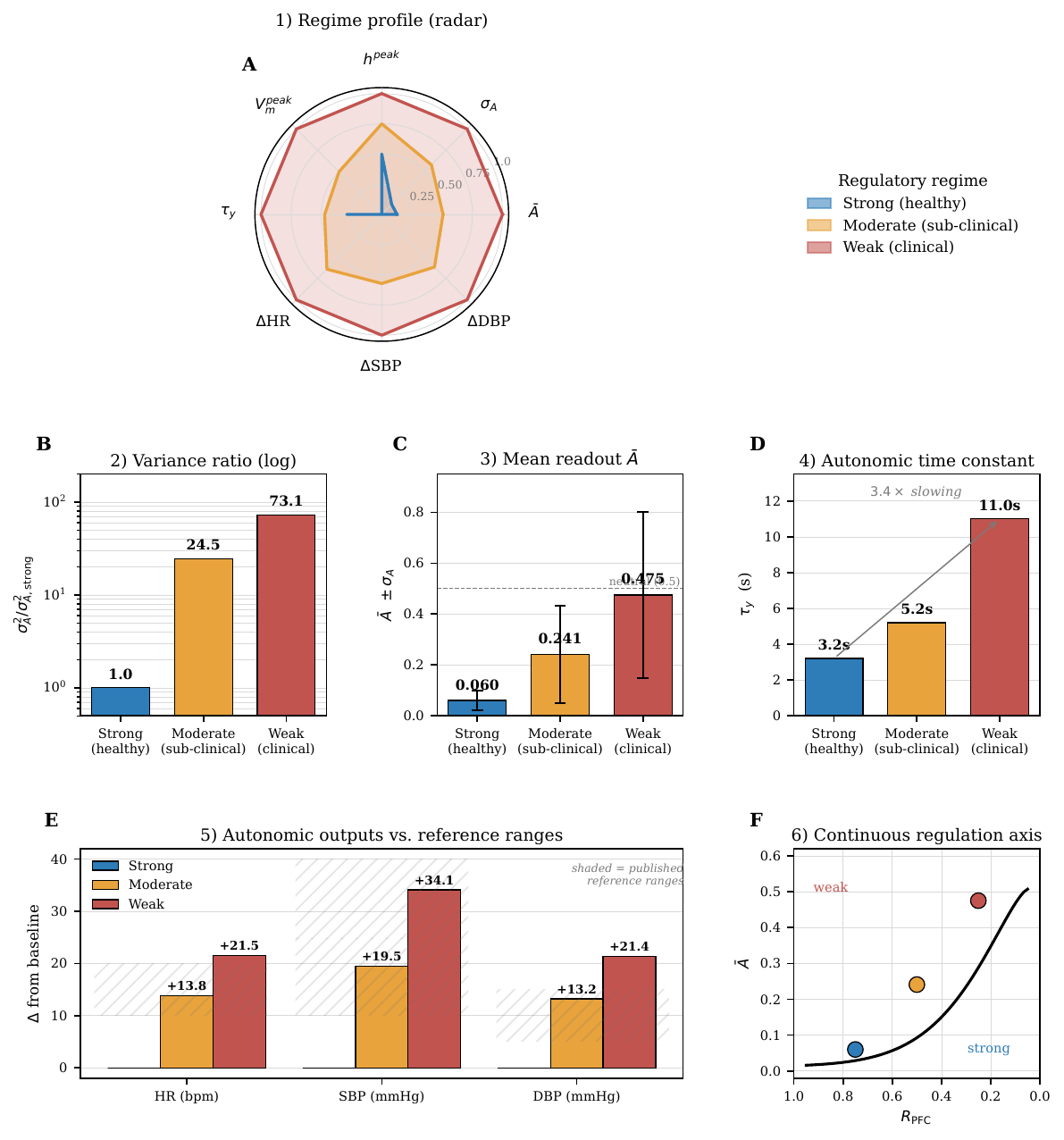}
   \caption{Overview of regime behavior across the
    anxiety/depression spectrum.}
  \label{fig:overview}
\end{figure}

Using the fixed biophysical parameters and simulation protocol of Sections II–III, varying only the cognitive‑affective modulators $(C, P, R_{\text{PFC}})$ (Table~\ref{tab:regimes}) produces three distinct regulatory regimes: strong (healthy), moderate (sub‑clinical), and weak (anxiety/depression) without altering the underlying amygdala–hypothalamus–cardiovascular architecture. Across these regimes, all key observables show consistent, monotone separation: amygdala threat acknowledgment becomes more frequent and more variable, hypothalamic activation becomes larger and more persistent, and cardiovascular excursions and recovery times increase. Figure~\ref{fig:overview} summarises this in a six‑panel dashboard (radar plot of regime profiles, headline amygdala statistics, autonomic time constants, cardiovascular outputs, and the projection of all three regimes onto one continuous regulation axis), previewing the quantitative findings detailed below.

\subsection{Sensory Drive and Slow Internal State} \label{sec:Sdrive and Sinternal state}
The stimulus protocol (Section~\ref{sec:Num Integ & Stimulus}) comprises five 400~ms rectangular pulses embedded in Ornstein-Uhlenbeck noise, which are transformed by the sensory transduction stage into a graded receptor potential $R(t)$. The chosen rise and decay time constants $\tau_{\text{r}} = 15$~ms and $\tau_{\text{d}} = 50$~ms ensure that each pulse remains clearly identifiable against the noise floor in $R(t)$, with peak receptor potentials of approximately 25--30~mV during pulses and small 3--5~mV fluctuations between them. This filtering preserves pulse identity while imposing an asymmetry between rapid depolarization and slower recovery, and the receptor cutoff frequencies are co-tuned with the membrane time constant $\tau_{\text{m}} = C_{\text{m}} / g_{\text{L}} = 10$~ms so that all three relevant timescales lie in the 3-10~Hz band, matching behaviorally relevant threat dynamics.

\begin{figure}[htbp]
  \centering
  \includegraphics[width=\linewidth]{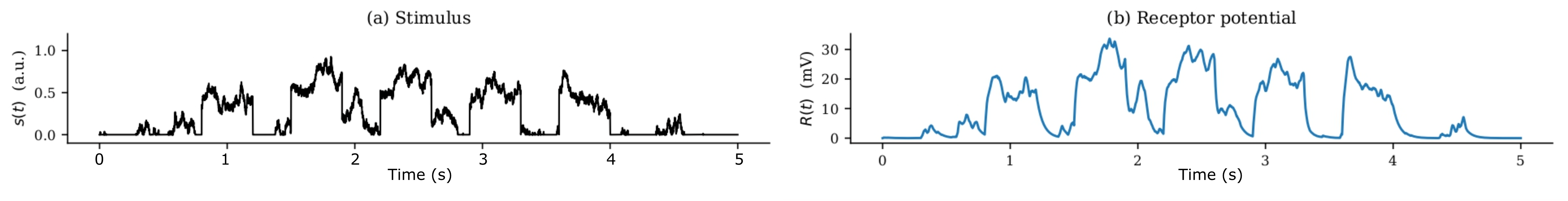}
   \caption{Slow internal state dynamics. (a) Stimulus $s(t)$. (b) Receptor potential $R(t)$.}
  \label{fig:stimulus}
\end{figure}

Figure~\ref{fig:stimulus} shows that the rectangular pulses are preserved in amplitude and timing while noise is smoothed; the asymmetry between rapid rises and slower decays in $R(t)$ then determines how transient inputs imprint on slower internal variables.
The slow internal state $h(t)$, introduced in Section~\ref{sec:internalstate} as a history-dependent adaptation variable, exhibits clear regime-dependent dynamics in response to the same $R(t)$. In the strong regime ($\tau_{\text{h}} = 40$~ms), $h(t)$ closely tracks the membrane potential and returns near its low equilibrium between pulses, resulting in limited cross-pulse accumulation. In the weak regime ($\tau_{\text{h}} = 220$~ms), $h(t)$ fails to fully reset between pulses and accumulates progressively over the train, approximately doubling its peak values relative to the strong regime. This behavior corresponds to a transition from rapid adaptation toward persistent sensitization, with $h(t)$ acting as a low-pass filter on $V_{\text{m}}(t)$ whose cutoff frequency shifts from around 4~Hz (strong) to below 1~Hz (weak).

\begin{figure}[htbp]
  \centering
  \includegraphics[width=\linewidth]{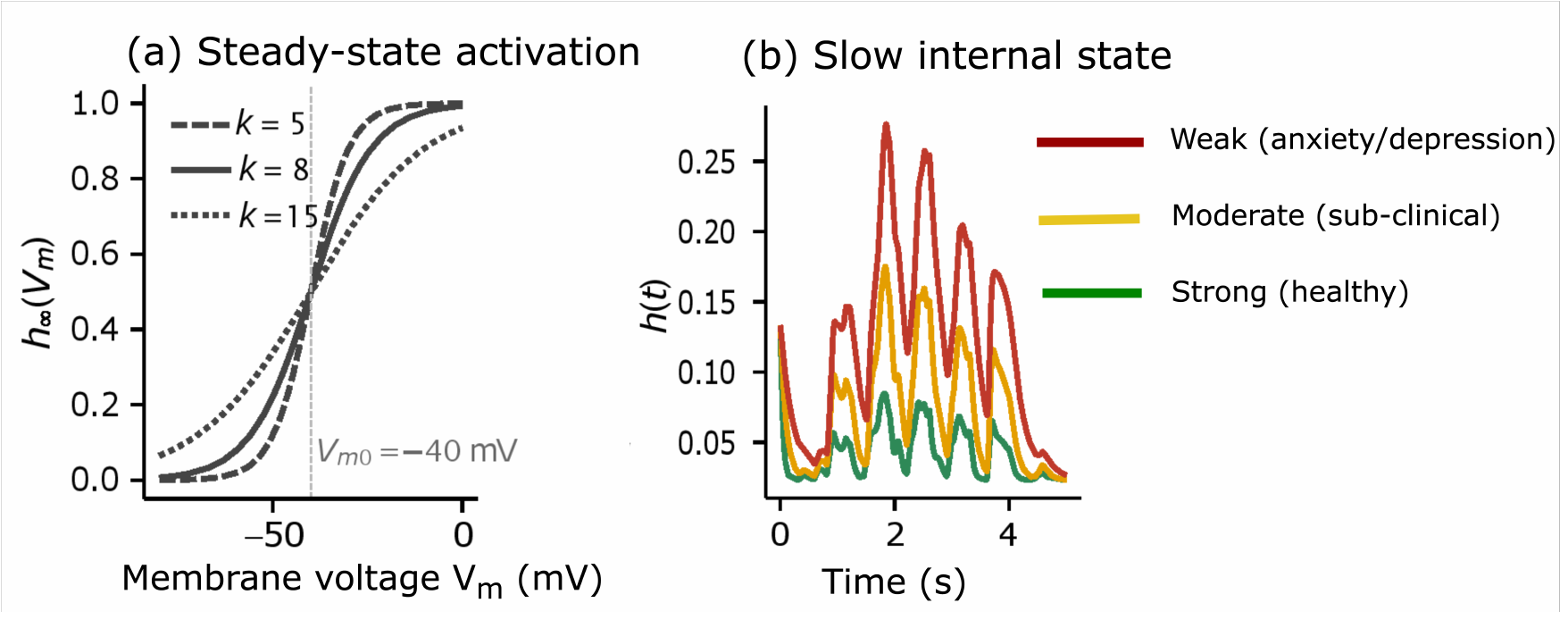}
   \caption{(a) Family of ~$h_\infty(V_m)$ curves.  (b) Time course of ~$h(t)$ for the three regimes.}
  \label{fig:slow internal}
\end{figure}

Figure~\ref{fig:slow internal}a shows $h_{\infty}(V_{\text{m}})$ for different slope factors, defining a slow manifold in the $(V_{\text{m}}, h)$ plane, and Figure~\ref{fig:slow internal}b the time courses of $h(t)$, contrasting tight tracking and rapid reset under strong regulation with cross-pulse build-up under weak regulation. These dynamics enter the effective inhibitory conductance $g_{\text{in,eff}}\ h(t)$ and so shape subsequent threat acknowledgment.

\subsection{Amygdala Membrane Dynamics and Threat Acknowledgment} \label{sec: Amygdala dynamics and threat ack}
At the amygdala level, regime differences appear as coordinated shifts in both membrane potential $V_{\text{m}}(t)$ and the adaptive threshold $\theta(t)$ defined by the modulators. Under strong regulation, excitatory drive is strongly gated by prefrontal control and inhibitory tone is high, yielding peak pulse-locked depolarizations near $-50$~mV and thresholds around $-50.25$~mV, so that $V_{\text{m}}(t)$ remains essentially sub-threshold throughout. As coping capacity decreases and perceived stress and prefrontal impairment increase, peak $V_{\text{m}}$ rises (e.g., to about $-45$ and $-40$~mV in moderate and weak regimes, respectively) while the adaptive threshold $\theta(t)$ shifts downward (e.g., to roughly $-53.50$ and $-56.75$~mV). Consequently, the peak distance $V_{\text{m}} - \theta$ changes from small negative in the strong regime to large positive in the weak regime, compressing the crossing distance by the combined effect of increased depolarization and lowered threshold.

\begin{figure}[htbp]
  \centering
  \includegraphics[width=\linewidth]{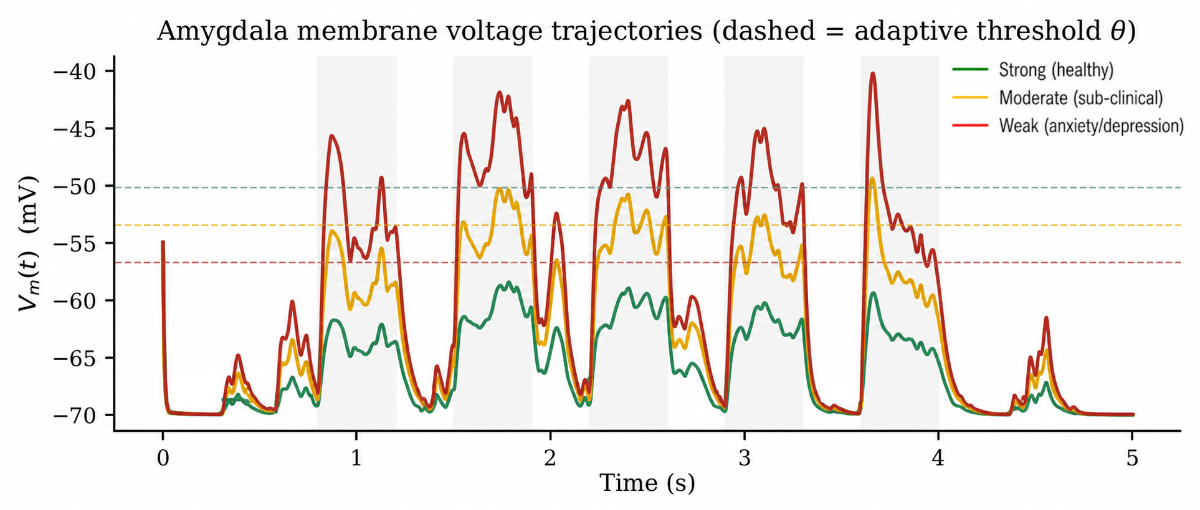}
   \caption{Membrane Voltage $V_{\text{m}}(t)$ and threshold $\theta(t)$.}
  \label{fig:membrane and threshold}
\end{figure}

Figure~\ref{fig:membrane and threshold} shows $V_{\text{m}}(t)$ with regime-specific thresholds overlaid: all regimes share similar baselines near the leak reversal potential between pulses, indicating that regime differences are not built into the resting state but emerge under stimulation. The strong regime never crosses threshold, the moderate regime crosses briefly during pulses, and the weak regime spends extended periods substantially above threshold.
The sigmoidal readout $A(u,t)$ converts these membrane-threshold relationships into graded threat-acknowledgment profiles. In the strong regime, $V_{\text{m}}(t) - \theta(t)$ remains negative, and $A$ stays below approximately $0.2$ throughout, corresponding to no-threat responses across the entire pulse train. In the moderate regime, $V_{\text{m}}(t) - \theta(t)$ crosses zero only transiently, and $A$ briefly exceeds $0.5$ during pulses before returning near zero between them. In the weak regime, $V_{\text{m}}(t) - \theta(t)$ is positive over most of the stimulation window, and $A$ remains above $0.5$ for extended intervals with peaks approaching $0.9$-$1$.

\begin{figure}[htbp]
  \centering
  \includegraphics[width=\linewidth]{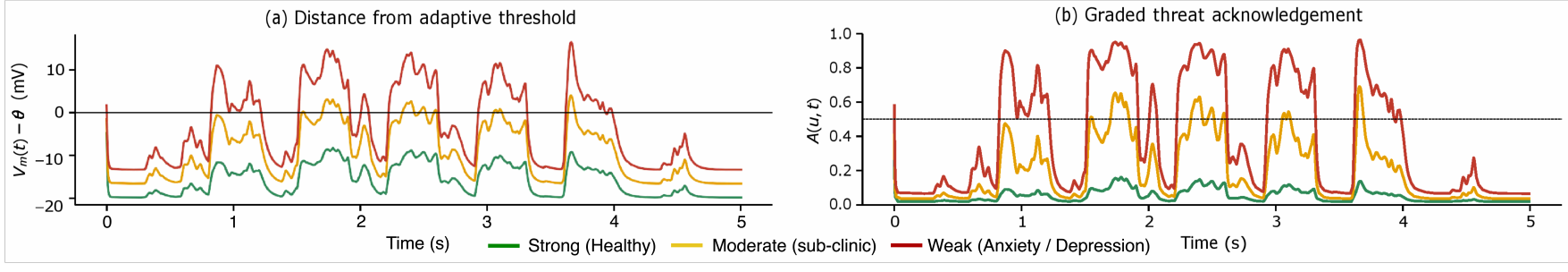}
   \caption{(a) Time course of $V_{\text{m}}(t)$-$\theta(t)$. (b) Sigmoid readout $A(u,t)$.}
  \label{fig:Timecourse and sigmoid}
\end{figure}

Figure~\ref{fig:Timecourse and sigmoid}a plots the time course of $V_{\text{m}}(t) - \theta(t)$, making the sign changes explicit across regimes. Figure~\ref{fig:Timecourse and sigmoid}b shows the corresponding $A(u,t)$, highlighting the qualitative shift from uniformly low responses (strong) to brief threshold crossings (moderate) and sustained high responses (weak) from the same input. Linearising $V_{\text{m}} - \theta$ with respect to the modulators shows that the partial derivatives with respect to $C, P, R_{\text{PFC}}$ have opposite signs in $V_{\text{m}}$ and $\theta$, so that their combined effect is additive on the decision variable, consistent with a collapse of signal-detection discriminability under impaired regulation.
The steady-state distribution of $A(u,t)$ makes this separation quantitative. In the strong regime, the distribution is narrowly concentrated near zero with negligible mass above $0.5$, indicating that threat acknowledgment is rare. In the moderate regime, the distribution broadens and shifts rightward, and the fraction of samples with $A > 0.5$ increases substantially. In the weak regime, the distribution becomes nearly flat across $[0,1]$, and the fraction of time spent with $A > 0.5$ increases by roughly two orders of magnitude compared with the strong regime, reflecting frequent and intense threat signaling. Mean and variance of $A$ both increase monotonically across regimes, and the variance ratio between weak and strong regimes reaches several tens-fold, with most of the amplification attributable to increased local sigmoid gain around its steepest region.
\begin{figure}[htbp]
  \centering
  \includegraphics[width=\linewidth]{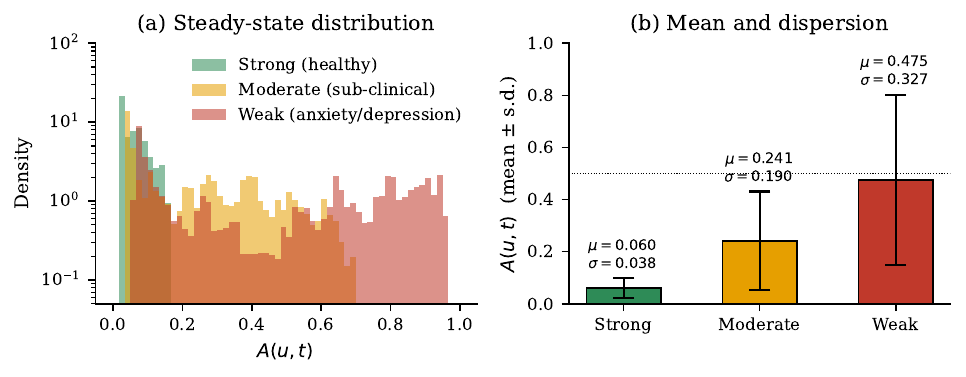}   \caption{(a) Steady state distribution.(b) Mean and variance of $A(u,t)$.}
  \label{fig:distribution and summary}
\end{figure}
Figure~\ref{fig:distribution and summary}a plots the density of $A$ on a log scale and Figure~\ref{fig:distribution and summary}b its mean and dispersion. Together, Figures~\ref{fig:membrane and threshold}-\ref{fig:distribution and summary} show that the modulators, through $V_{\text{m}}$ and $\theta(t)$, induce robust, graded separation in both time-resolved and distributional measures of threat acknowledgment.

\subsection{Hypothalamic Integration and Autonomic Response} \label{sec: Hypo and response}

Once the input signal is received from the amygdala, the hypothalamic integrator and baroreflex loop transform the graded threat signal into autonomic drive and cardiovascular responses. The suprathreshold input $y(t)$ is identically zero in the strong regime because $A(u,t)$ rarely exceeds its midline, exhibits brief positive bursts time-locked to pulses in the moderate regime, and remains elevated for extended periods in the weak regime. This pattern mirrors the amygdala output while applying an additional nonlinearity that ignores sub-threshold threat signals.
\begin{figure}[htbp]
  \centering
  \includegraphics[width=\linewidth]{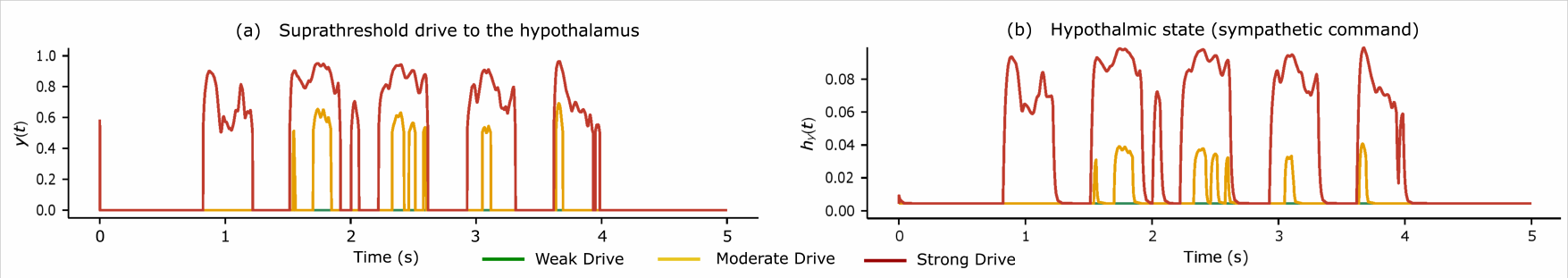}
   \caption{(a) Output $y(t)$ from amygdala to hypothalamus.(b) Hypothalamic $h_{\text{y}}(t)$ accumulation under different regime.}
  \label{fig:output & hypo}
\end{figure}
Figure~\ref{fig:output & hypo}a shows $y(t)$ and Figure~\ref{fig:output & hypo}b the hypothalamic state $h_{\text{y}}(t)$. In the strong regime $h_{\text{y}}(t)$ stays near its low baseline; in the moderate regime it rises during pulses and decays between them; in the weak regime elevated drive with reduced baroreflex feedback (smaller $\beta$ and $\gamma$) gives higher peaks and slower decay, producing activations several-fold larger and more prolonged than in the strong regime. Linearisation yields autonomic time constants $\tau_{\text{y}}$ rising from a few seconds (strong) to roughly an order of magnitude larger (weak), consistent with delayed cardiovascular recovery under chronic stress.
\begin{figure}[htbp]
  \centering
  \includegraphics[width=\linewidth]{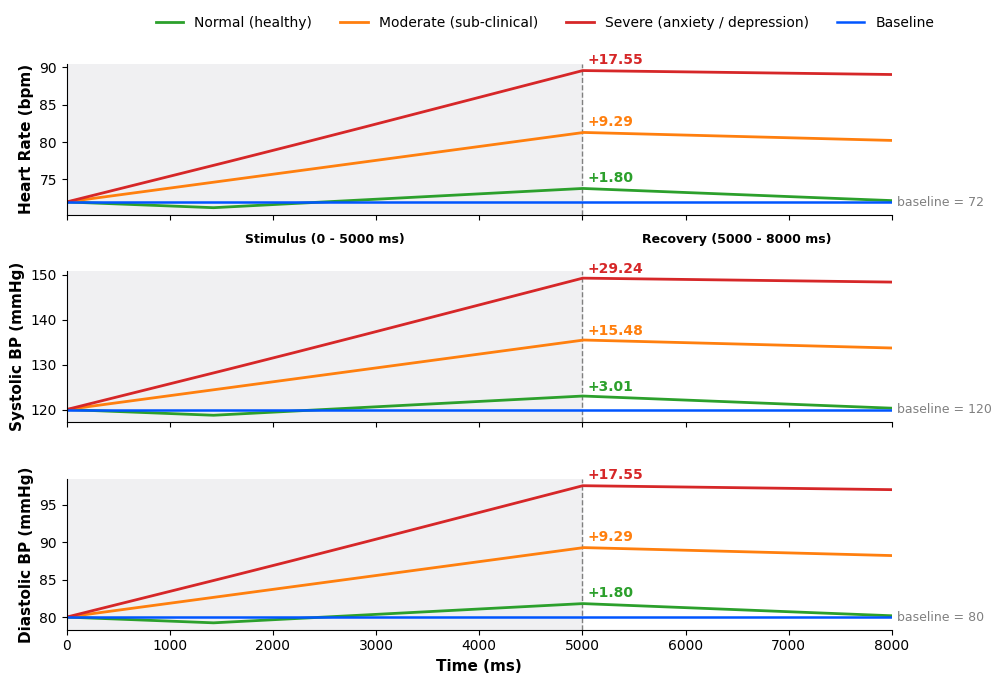}
   \caption{HR/SBP/DBP response and recovery.}
  \label{fig:cardio output}
\end{figure}
These dynamics translate into heart-rate and blood-pressure changes via the linear haemodynamic relations of section~\ref{sec:threatacknow}. In the normal, moderate, and severe regimes, the model produces peak HR, SBP, and DBP increases within meta-analytic acute-stress ranges, with recovery slowing monotonically as regulation weakens (fig.~\ref{fig:cardio output}). The normal (healthy) regime shows very small excursions and a rapid return toward baseline due to healthy homeostatic regulation. The moderate (sub-clinical) regime produces larger excursions and intermediate recovery. Finally, the severe (anxiety / depression) regime exhibits the maximum excursions and minimal recovery within the 8000 ms observation window. This combines heightened reactivity with impaired homeostatic return, while staying within empirically observed ranges.

\subsection{Sensitivity to Regulatory Modulators} \label{sec: sensitivity and regulatory}

To map behaviour along the regulatory axis, we swept $R_{\text{PFC}}$ and $P$ with other parameters fixed at regime-specific values. Varying $R_{\text{PFC}}$ from near-intact to severely impaired control yields smooth, monotone changes in mean amygdala output, log dispersion, and mean hypothalamic activation, without discontinuities or bifurcations, so the three regimes occupy positions on a continuous regulation curve rather than separate dynamical phases.
Complementary sweeps of $P$ at fixed $C$ and $R_{\text{PFC}}$ show HR, SBP, and DBP ranges expanding with $P$, their upper bounds crossing clinical hypertension thresholds only under weak regulation for identical input: strong regulation stays within healthy limits, moderate approaches the upper normative range, and weak exceeds clinical-risk thresholds. This supports a dimensional view of anxiety and depression as graded shifts in regulatory state affecting both central and autonomic responses, in line with transdiagnostic perspectives.
\begin{figure}[htbp]
  \centering
  \includegraphics[width=\linewidth]{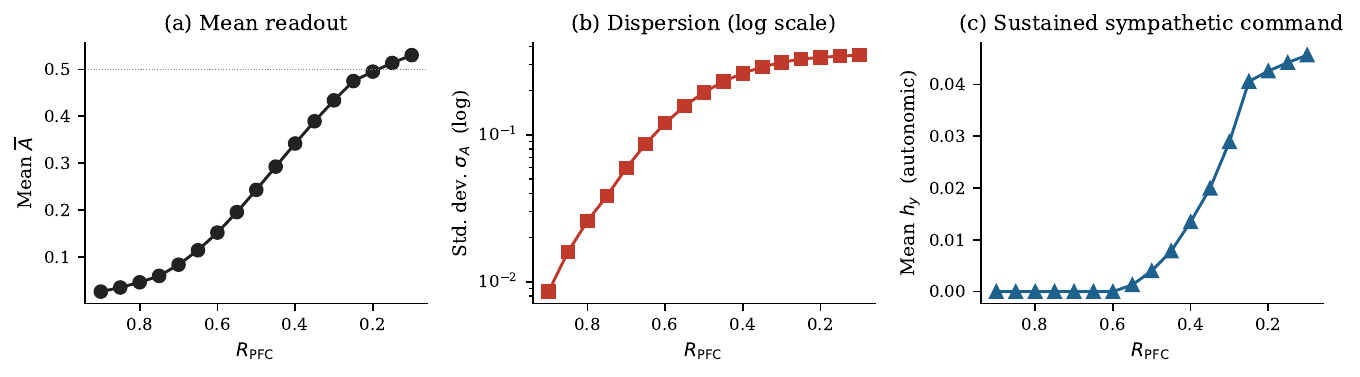}
\caption{Parametric sweep over $R_{\text{PFC}}$.}
\label{fig:parametric sweep}
\end{figure}
Figure~\ref{fig:parametric sweep} plots mean threat acknowledgment, log dispersion, and mean hypothalamic activation against $R_{\text{PFC}}$. The absence of abrupt changes and the smooth interpolation between the three labelled regimes reinforce the reading of the model as a continuous regulatory landscape rather than discrete categories.

\section{Validation and Empirical Evaluation} \label{sec:validation}

The Results in Section~\ref{sec:result} showed that varying the modulators $(C, P, R_{\text{PFC}})$ in a fixed amygdala--hypothalamus--cardiovascular architecture produces robust separation of strong, moderate, and weak regulatory regimes in both neural and autonomic variables. This section evaluates the reliability and physiological credibility of those findings through internal robustness analyses, component ablations, comparison with published cardiovascular ranges, and empirical validation on independent datasets.

\subsection{Robustness to Noise and Parameter Perturbation} \label{sec:robust to noise}
\begin{figure}[htbp]
  \centering
  \includegraphics[width=\linewidth]{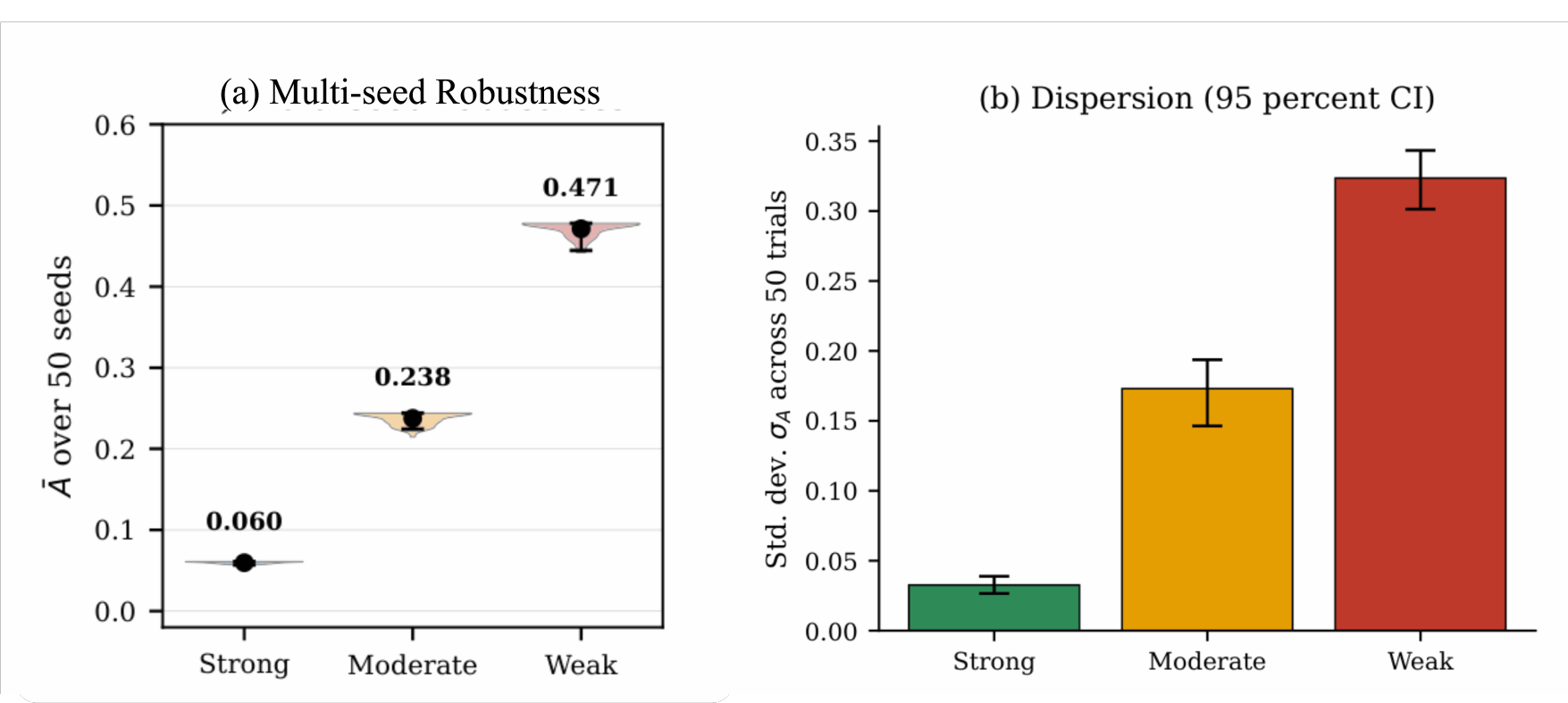}
\caption{(a) Mean$A$ with $95\%$ confidence intervals over $50$ seeds for each regime. (b) Variance with confidence intervals.}
\label{fig:robustness}
\end{figure}

To test robustness to stochastic variability, the full protocol (Section~\ref{sec:simulationprtocol}) was repeated for 50 independent Ornstein-Uhlenbeck noise realisations per regime (Fig.~\ref{fig:robustness}a) with parameters fixed at nominal values. Regime-specific 95\% confidence intervals for the mean output $A$ stayed non-overlapping, with inter-regime gaps tens of times larger than within-regime variability. Dispersion $\sigma_A$ (Fig.~\ref{fig:robustness}b) showed the same non-overlapping pattern, confirming that the rise in variability from strong to weak regulation is not a noise artefact.

\begin{figure}[htbp]
  \centering
  \includegraphics[width=\linewidth]{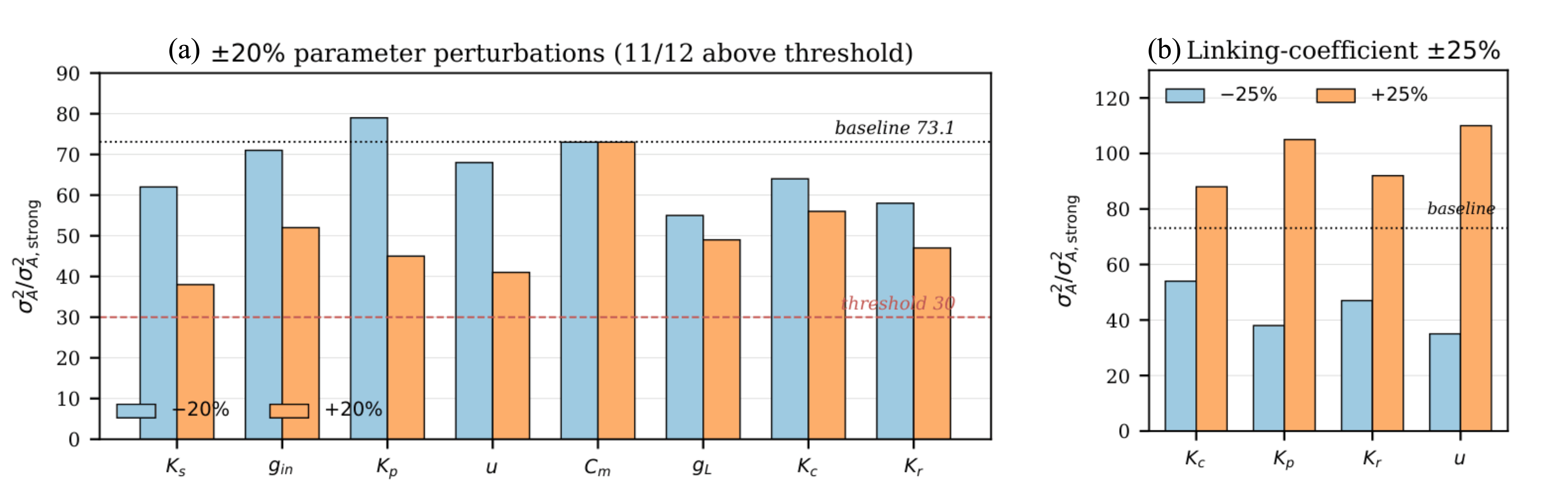}
\caption{(a) Variance ratios under ~$\pm 20\%$ perturbations of core. (b) Regime means/variance ratios under ~$\pm 25\%$ perturbations of linking coefficients.}
\label{fig:perturbation}
\end{figure}

Structural robustness was further assessed by perturbing key parameters by $\pm 20\%$ (Fig.~\ref{fig:perturbation}a). For six core parameters ($K_{\text{s}}, g_{\text{in}}, K_{\text{p}}, u, C_{\text{m}}, g_{\text{L}}$), the weak/strong variance ratio of $A$ stayed above unity in all 12 perturbations and above $30$ in 11 of 12, with the ordering $\text{strong} < \text{moderate} < \text{weak}$ preserved. Perturbing the linking coefficients ($K_{\text{c}}, K_{\text{p}}, K_{\text{r}}, u$) by $\pm 25\%$ likewise preserved ordering and kept every statistic within a factor of two (Fig.~\ref{fig:perturbation}b). The regime separation of Section~\ref{sec:result} is therefore structurally stable, not an artefact of a narrow parameter choice or single seed.

\subsection{Component Necessity: Ablation of Slow Gating and Adaptive Threshold} \label{sec:ablation}

\begin{figure}[htbp]
  \centering
  \includegraphics[width=\linewidth]{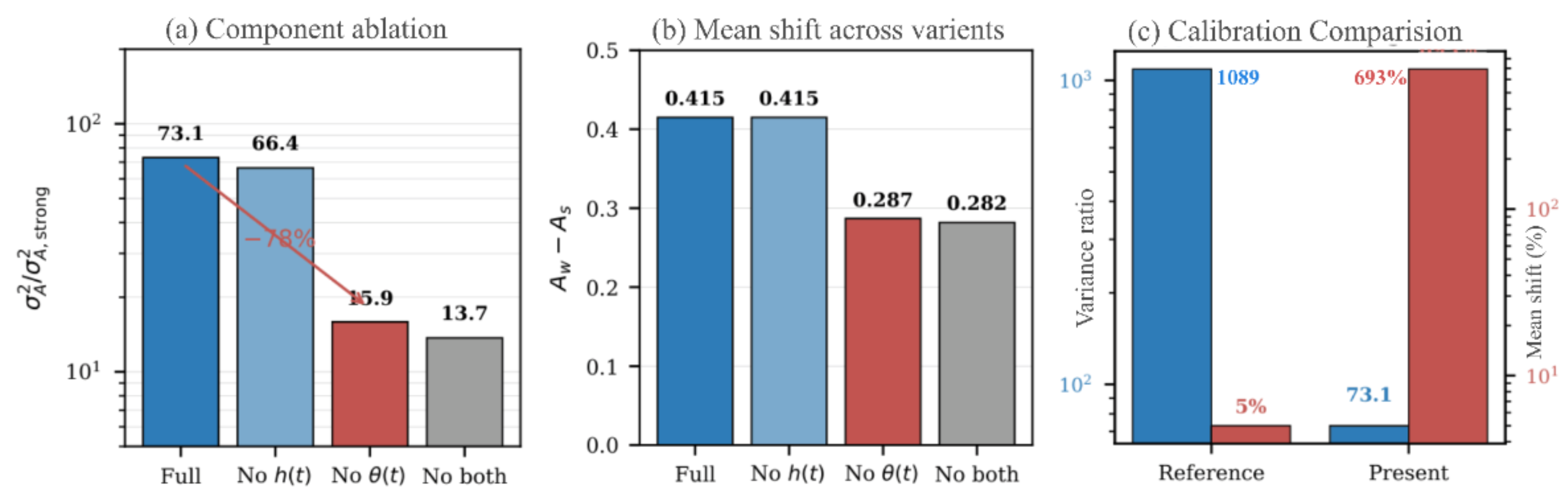}
\caption{(a) Variance ratio (log scale) for full, no-$h$, no-$\theta$, and joint-ablation variants. (b) Mean shift across variants. (c) Variance and mean shift under both calibrations.}
\label{fig:ablation}
\end{figure}

To localise which components of the architecture are necessary for regime separation, four variants (figure~\ref{fig:ablation}a)  were compared under identical stimulation and modulator settings, 1. Full model, 2. No slow gating with $h(t)$ clamped to $h_{\infty}(V_{\text{m}})$, 3. No adaptive threshold with $\theta(t)$ fixed at $\theta_0$, and 4. Joint ablation with both slow gating and adaptive threshold removed. The full model exhibited a large variance ratio of $A$ between weak and strong regimes (e.g, $\approx 73$) and substantial mean shift, as reported in Section~\ref{sec:result}. Removing slow gating alone produced little change in these metrics: the variance ratio and mean shift remained close to the full-model values, indicating that $h(t)$ primarily amplifies existing differences rather than generating them. In contrast, removing the adaptive threshold reduced the variance ratio to around $16$ and decreased the mean shift in figure~\ref{fig:ablation}b, although the regime ordering $\text{strong} < \text{moderate} < \text{weak}$ persisted. Joint ablation of both mechanisms produced only a modest additional reduction relative to removing the threshold alone and again preserved regime ordering.
These results identify the adaptive, modulator-dependent threshold as the principal load-bearing component for the quantitative separation of regimes, with slow gating functioning mainly as a multiplicative amplifier on top of the threshold mechanism. The fact that ordering persists even under joint ablation underscores that the overall architecture not any single parameter is responsible for the directional pattern of increased threat acknowledgement under weak regulation.

\subsection{Physiological Plausibility and Calibration Sensitivity} \label{sec:physilogy valid}

The autonomic outputs in Section~\ref{sec: Hypo and response} were generated without fitting to any dataset, so plausibility was assessed against meta-analytic acute-stress studies \cite{sadoun2025,blasi2005,hyndman1996}. Acute psychological stress typically raises HR by $10$-$20$~bpm and SBP by $10$-$40$~mmHg, with corresponding DBP changes and delayed recovery under chronic stress or low resilience. In the moderate and weak regimes the model produced peak increments of about $+13$-$22$~bpm (HR) and $+20$-$35$~mmHg (SBP) (Fig.~\ref{fig:cardio output}), with time constants $\tau_{\text{y}}$ from a few seconds (strong) to about ten seconds (weak). These values fall within reported ranges, so the outputs are quantitatively plausible without per-signal tuning \cite{sadoun2025,blasi2005}.

To examine how calibration affects absolute magnitudes, two parameterisations of the same equations were compared figure~\ref{fig:ablation}c. In a tight-tracking calibration, the amygdala output $A$ was constrained to operate near the sigmoid inflection point, resulting in extremely high variance ratios (e.g., weak/strong $\approx 1000$) but small mean shifts ($\approx 5\%$), with regime points tightly clustered around $A \approx 0.5$. In the present graded-excursion calibration, $A$ spans a broader range, producing lower but still substantial variance ratios ($\approx 70$) and much larger mean shifts ($\approx 700\%$), with regime means distributed across $[0,1]$. In both calibrations, however, the ordering $\text{strong} < \text{moderate} < \text{weak}$ was preserved for both mean $A$ and dispersion, showing that the qualitative pattern and relative regime differences are calibration-invariant, while the absolute magnitudes depend on operating point.

\subsection{External Empirical Validation} \label{sec:external}
External validity of the proposed framework was evaluated using three independent datasets extending across mental-health monitoring, occupational stress, and stroke rehabilitation, each probing a different aspect of the model’s cardiovascular stimulus-response and recovery behavior.

The first dataset was a large-scale mental-health monitoring collection of 52,585 logistics-driver records, of which 15,723 with documented mental-health histories were retained \cite{truckdrivers2021}; each contained HR, SBP, DBP, and psychological context. Resilience and cognitive-load indicators served as proxies for coping capacity and perceived stress, and the cardiovascular variables mapped directly to model outputs. Records were stratified into resilience-based groups approximating moderate and severe stress, with quartile statistics giving baseline inputs per regime. Simulated peak HR, SBP, and DBP (Table~\ref{tab:model-data validation}) matched observed peaks to a mean absolute percentage error of about 3.7\% (typically 2-5\% per variable), so the model reproduces real-world responses across stress and coping levels without architectural changes.

\begin{table}[htbp]
\caption{Model-Dataset parameter comparison}
\label{tab:model-data validation}
  \centering
  \footnotesize 
  \setlength{\tabcolsep}{4pt} 
\begin{tabularx}{\columnwidth}{>{\raggedright\arraybackslash}Xcccc}
\toprule
Parameter & \shortstack{Moderate\\(Model/Data)} & Error (\%) & \shortstack{Severe\\(Model/Data)} & Error (\%) \\
\midrule
HR        & 101.45 / 104.33       & 2.76       & 111.18 / 105.50     & 5.38       \\
SBP      & 151.84 / 155.50       & 2.35       & 168.00 / 160.00     & 5.00       \\
DBP       & 100.56 / 104.34       & 3.66       & 108.36 / 105.41     & 2.80       \\
\bottomrule
\end{tabularx}
\end{table}

\begin{figure}[htbp]
  \centering
  \includegraphics[width=\linewidth]{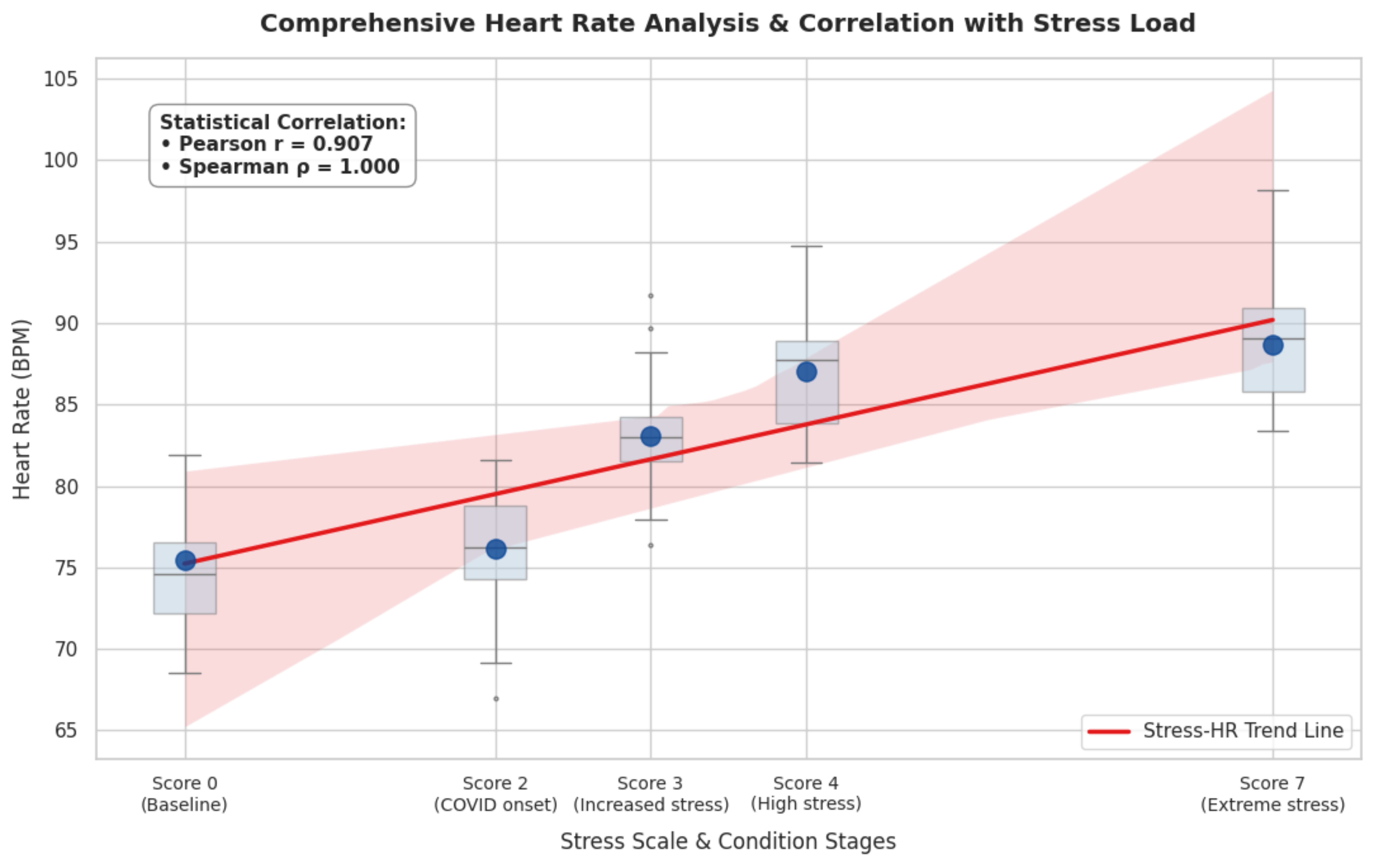}
\caption{Comprehensive heart rate alterations across advancing stress loads}
\label{fig:comprehensive stress loads}
\end{figure}

The second dataset comprised longitudinal HR recordings from a healthcare worker across five ordered occupational-stress states during the COVID-19 pandemic, from baseline workload to prolonged exposure to high-risk environments \cite{hosseini2022}, which map to progressive increases in $P$ with other modulators fixed. Median HR and interquartile range shifted monotonically upward across states (Fig.~\ref{fig:comprehensive stress loads}), with a strong positive linear relationship between state-level stress and median HR (Pearson $r=0.907$) and perfect monotonic ordering (Spearman $p=1.0$). Cardiovascular activation thus rose systematically with cumulative stress burden, consistent with $P$ acting as a continuous modulator that produces graded, sustained elevations rather than isolated peaks.

For the stroke-rehabilitation cohort, between-state hemodynamic differences across functional states (rest, sit-to-stand, walking, sleep) were quantified using a Linear Mixed-Effects Model to handle the unbalanced repeated-measures design across 90 survivors \cite{novak2010}. Individual patient identifiers were fitted as random intercepts, estimated using Restricted Maximum Likelihood (REML), allowing subject-level baseline variance to be separated from systematic effects of state. Summarized in Table~\ref{tab:lme-regression} , the fixed-effect coefficients showed highly significant, bidirectional shifts relative to resting baseline: ambulatory and orthostatic conditions (especially Walk Pre-Lunch and Sit-to-Stand) produced significant positive changes in $\Delta\text{SBP}$, $\Delta\text{DBP}$, and $\Delta\text{HR}$, whereas Sleep produced large, uniformly negative changes in all three measures (all $p<.001$). These graded, statistically distinguishable trends match the model’s stimulus–response and recovery mechanisms, in which functional stressors increase central amygdala–hypothalamus drive and downstream vasoconstriction and heart rate modulation, while low-stimulus sleep windows reduce drive and permit homeostatic recovery via enhanced parasympathetic feedback.

\begin{table}[htbp]
\caption{Linear Mixed-Effects Model Regression Results for Hemodynamic Changes}
\label{tab:lme-regression}
  \centering
  \footnotesize 
  \setlength{\tabcolsep}{4pt}
\begin{tabularx}{\columnwidth}{>{\raggedright\arraybackslash}Xccc}
\toprule
Predictor Variable & \shortstack{$\Delta$ Systolic BP\\(z-score)} & \shortstack{$\Delta$ Diastolic BP\\(z-score)} & \shortstack{$\Delta$ Heart Rate\\(z-score)} \\
\midrule
Intercept (Resting Baseline\tnote{*}) & 1.26 (1.26) & 0.29 (0.51) & 0.87 (1.71) \\
Sit to Stand (AM) & 1.30 (0.98) & 2.64 (3.63)$^{***}$ & 3.65 (5.70)$^{***}$ \\
Sit to Stand (PM) & 2.44 (1.84) & 3.17 (4.33)$^{***}$ & 1.67 (2.60)$^{**}$ \\
Sleep & -5.09 (-3.80)$^{***}$ & -4.88 (-6.60)$^{***}$ & -4.93 (-7.55)$^{***}$ \\
Walk (PM) & 0.10 (0.07) & 0.09 (0.13) & 1.48 (2.32)$^{*}$ \\
Walk (Pre-Lunch) & 3.83 (2.90)$^{**}$ & 3.60 (4.93)$^{***}$ & 2.60 (4.04)$^{***}$ \\
\midrule
Patient Random Intercept ($\sigma^2$) & 12.67 & 4.24 & 4.50 \\
Residual Variance ($\sigma^2$) & 75.87 & 23.17 & 17.96 \\
\bottomrule
\multicolumn{4}{l}{\rule{0pt}{1.2em}$^{*}\ p < .05$, $^{**}\ p < .01$, $^{***}\ p < .001$.} \\
\multicolumn{4}{l}{Values are expressed as Coefficient (z-statistic).} \\
\multicolumn{4}{p{\columnwidth}}{$^{*}$Note: Fixed effects are reported relative to the quiet Seated/Resting baseline activity.} \\
\end{tabularx}
\end{table}

Taken together, the three datasets provide convergent empirical support and demonstrate that the proposed framework (i) reproduces real‑world cardiovascular magnitudes under inferred combinations of stress and coping, (ii) predicts monotonic shifts in heart‑rate distributions with increasing stress burden, and (iii) captures statistically and physiologically coherent changes in cardiovascular response and recovery across distinct functional states. This convergent evidence supports the framework that the stimulus-amygdala–hypothalamus–cardiovascular architecture captures both the directional and approximate quantitative characteristics of stress‑related autonomic regulation observed in practice.

\subsection{Interpretation: Model Regimes and Stress Physiology} \label{sec:interpretation}
Read as a closed-loop controller, the three regimes correspond to different settings of a single feedback system rather than to separate mechanisms. Strong regulation restrains cardiovascular perturbation and accelerates baseline recovery after a standardised stressor, whereas weak regulation produces elevated HR and BP peaks with delayed recovery under identical input. This pattern matches autonomic-physiology meta-analyses, in which chronic stress and affective disorders are associated with larger, more prolonged cardiovascular reactivity and slower, resilience-linked recovery \cite{vaccarino2024,cheng2022,tomasi2024}. The model therefore provides a structural account, rather than a black-box predictor, of how psychological and neurocircuit changes alter amygdala excitability to shape the autonomic profiles recorded by wearables \cite{sadoun2025,booth2022,choi2024}.

\subsection{Clinical and Translational Applications} \label{sec:clinical}
By mapping psychometric and regulatory scores directly to cardiovascular trajectories, the framework serves as both an interpretive tool and a simulation testbed. It lets variation in wearable telemetry be read as movement along a continuous regulatory axis, for example reduced prefrontal control or increased stress load, rather than as isolated statistical shifts, so that a prolonged post-stress tachycardia maps to a state nearer the weak-regulation regime. As a simulation testbed it supports in-silico parameter isolation, allowing targeted behavioural or neuromodulatory interventions to be evaluated for their effect on threat-acknowledgment dynamics and autonomic recovery \cite{bjornsson2019,laubenbacher2022,sadoun2025}, and can distinguish whether a biomarker profile reflects excessive stress drive or insufficient prefrontal gating. Embedded in a wearable pipeline, longitudinal tracking of parameter drift toward weaker regulation provides an early-warning metric aligned with digital-phenotyping and digital-psychiatry frameworks \cite{choi2024,booth2022,zhao2025,torous2021,insel2023}.

\subsection{Assumptions, Limitations, and Scope} \label{sec:limits}
The framework is an explanatory model bounded by several simplifications. The amygdala is represented as a single compartment with an implicit spiking mechanism, prioritising subthreshold integration and graded readouts over spike timing, dendritic morphology, and subnuclei diversity \cite{janak2015,tovote2015}. The hypothalamic integrator and baroreflex loop are phenomenological, capturing macro-level feedback without resolving individual brainstem or hypothalamic nuclei \cite{schaeuble2022,sadoun2025,blasi2005,hyndman1996}. Validation relies on cohort-level aggregate data rather than continuous beat-to-beat series, so structural behaviours such as regime ordering and directional trends are better supported than personalised predictive claims \cite{hosseini2022,novak2010}. The model also excludes the hypothalamic-pituitary-adrenal (HPA) axis and inflammatory signalling, assuming that short-term cardiovascular fluctuations are primarily neurocircuit-driven \cite{matsubara2021,matsubara2019,lindsey2009,cline2004}, and an operational pipeline would further require real-time acquisition and inverse-estimation algorithms validated under clinical AI guidelines \cite{torous2021,insel2023}. Within these bounds, the model shows how psychometric modulators can be embedded in a biophysical substrate to generate realistic autonomic profiles, establishing a basis for future patient-specific parameterisation.

\section{Conclusion}

We developed a mechanistic model of the amygdala–hypothalamus–cardiovascular network by embedding psychometrically grounded modulators into a conductance-based neural framework and propagating their effects to heart-rate and blood-pressure dynamics. Varying coping capacity, perceived stress load, and prefrontal regulatory strength within a compact nine-equation structure, we showed that healthy-like, subclinical, and anxiety/depression-like regimes emerge as distinct operating points of a single closed-loop system. Internal robustness analyses and targeted ablations established that these regime differences rest on an adaptive-threshold architecture, and external validation against three independent datasets showed that the autonomic outputs align with observed stress-related cardiovascular patterns. By bridging cellular-level dynamics, clinically interpretable modulators, and wearable-level signals, the framework supports both the interpretation of digital-phenotyping data and the generation of mechanistic hypotheses about how specific regulatory adjustments alter stress responses, providing a scalable foundation for personalized, interpretable monitoring of anxiety- and depression-related autonomic regulations.
Future work will expand the framework's output layer to model respiratory rate dynamics alongside heart rate and blood pressure across distinct regimes. Furthermore, we will formalize these regimes as proxy states analogous to a flip-flop architecture to capture macroscopic state-switching behaviors while respecting the brain's graded, probabilistic dynamics. These expanded dynamics will ultimately be leveraged to pursue individual-level parameter inference and embed the model within digital-twin pipelines for mental-health decision support.


%

\section*{Acknowledgment}

The authors would like to thank the Walton Institute for Information and Communication Systems Science, Southeast Technological University, Ireland, for supporting this research under Grant WD\_2023\_46\_WSCH.

\ifCLASSOPTIONcaptionsoff
  \newpage
\fi

\end{document}